\documentstyle[psfig]{cupconf}

\ifoldfss
\else
  \ifnfssone
    \newmathalphabet{\mathit}
      \addtoversion{normal}{\mathit}{cmr}{m}{it}
      \addtoversion{bold}{\mathit}{cmr}{bx}{it}
    \newmathalphabet{\mathcal}
      \addtoversion{normal}{\mathcal}{cmsy}{m}{n}
    \else
    \ifnfsstwo
    \fi
  \fi
\fi

\newcommand{\bea}{\begin{eqnarray}}
\newcommand{\eea}{\end{eqnarray}}
\newcommand{\be}{\begin{equation}}
\newcommand{\ee}{\end{equation}}
\newcommand{\bc}{\begin{center}}
\newcommand{\ec}{\end{center}}
\newcommand{\ben}{\begin{enumerate}}
\newcommand{\een}{\end{enumerate}}
\newcommand{\bi}{\begin{itemize}}
\newcommand{\ei}{\end{itemize}}
\newcommand{\bmi}[1]{\begin{minipage}{#1 cm}}
\newcommand{\emi}{\end{minipage}}
\newcommand{\itRahmen}[2]{\begin{center}
\fbox{
  \parbox{#1 cm}{\it #2}}
\end{center}}

\newcommand{\ebox}[1]{\fbox{$\displaystyle #1 $}}
\def\eck#1{\left\lbrack #1 \right\rbrack}

\def\rund#1{\left( #1 \right)}
\def\abs#1{\left\vert #1 \right\vert}

\def\ave#1{\left\langle #1 \right\rangle}

\def\Re{{\cal R}\hbox{e}}
\def\Im{{\cal I}\hbox{m}}
\def\A{{\cal A}}

\def\D{{\cal D}}

\def\P{{\cal P}}

\def\d{{\rm d}}

\def\eps{{\epsilon}}

\def\arcsecf {\hbox{$.\!\!^{\prime\prime}$}}

\def\vp{\varphi}
\def\vt{{\vartheta}}
\def\kB{{k_{\rm B}}}

\def\Real{{\rm I\mathchoice{\kern-0.70mm}{\kern-0.70mm}{\kern-0.65mm}%
  {\kern-0.50mm}R}}
\def\C{\rm C\kern-.42em\vrule width.03em height.58em depth-.02em
       \kern.4em}
\font \bolditalics = cmmib10
\def\bx#1{\leavevmode\thinspace\hbox{\vrule\vtop{\vbox{\hrule\kern1pt
        \hbox{\vphantom{\tt/}\thinspace{\bf#1}\thinspace}}
      \kern1pt\hrule}\vrule}\thinspace}

\def \vc #1{{\textfont1=\bolditalics \hbox{$\bf#1$}}}
{\catcode`\@=11
\gdef\SchlangeUnter#1#2{\lower2pt\vbox{\baselineskip 0pt \lineskip0pt
  \ialign{$\m@th#1\hfil##\hfil$\crcr#2\crcr\sim\crcr}}}
}
\def\gtrsim{\mathrel{\mathpalette\SchlangeUnter>}}
\def\lesssim{\mathrel{\mathpalette\SchlangeUnter<}}

\def\ueber#1#2{{\setbox0=\hbox{$#1$}%
  \setbox1=\hbox to\wd0{\hss$\scriptscriptstyle #2$\hss}%
  \offinterlineskip
  \vbox{\box1\kern0.4mm\box0}}{}}
\def\llabel#1{\label{sc:#1}}
\def\elabel#1{\label{eq:#1}}
\def\flabel#1{\label{fig:#1}}
\def\Real{\mbox{Re}}      
\def\hexnumber#1{\ifcase#1 0\or1\or2\or3\or4\or5\or6\or7\or8\or9\or
 A\or B\or C\or D\or E\or F\fi }

\makeatletter
\ifx\CUP@mtlplain@loaded\undefined
\else
\fi
\makeatother
 \makeatletter
 \ifx\CUP@mtlplain@loaded\undefined
   \font\tenbmi=cmmib10 at 10pt
   \font\sevenbmi=cmmib10 at 7pt
   \font\fivebmi=cmmib10 at 5pt

   \newfam\bmifam
   \textfont\bmifam=\tenbmi
   \scriptfont\bmifam=\sevenbmi
   \scriptscriptfont\bmifam=\fivebmi
   \def\bmi{\fam\bmifam\tenbmi}
 \fi
 \makeatother
\ifnfsstwo

\fi
\ifnfssone

\fi
\ifoldfss

\fi

\mathchardef\varLambda="0103

\makeatletter
\ifx\CUP@mtlplain@loaded\undefined
\else
\fi
\makeatother
\makeatletter
\ifx\CUP@mtlplain@loaded\undefined
  \font\tenbms=cmbsy10
  \font\sevenbms=cmbsy10 at 7pt
  \font\fivebms=cmbsy10 at 5pt
  \newfam\bmsfam
  \textfont\bmsfam=\tenbms
  \scriptfont\bmsfam=\sevenbms
  \scriptscriptfont\bmsfam=\fivebms

  \edef\bsy@{\hexnumber\bmsfam}
  \mathchardef\bnabla="0\bsy@72
\fi
\makeatother
\title[Gravitational lensing as a probe of structure]{Gravitational
  lensing as a probe of structure}

\author[Peter Schneider]%
{P\ls E\ls T\ls E\ls R\ns
S\ls C\ls H\ls N\ls E\ls I\ls D\ls E\ls R$^1$}

\affiliation{$^1$Institut f\"ur Astrophysik und Extraterrestrische
Forschung, Universit\"at Bonn, Germany}

\begin{document}
\ifnfssone
\else
  \ifnfsstwo
  \else
    \ifoldfss
      \let\mathcal\cal
      \let\mathrm\rm
      \let\mathsf\sf
    \fi
  \fi
\fi

\maketitle

\begin{abstract}
Gravitational lensing has become one of the most interesting tools to
study the mass distribution in the Universe. Since gravitational light
deflection is independent of the nature and state of the matter, it is
ideally suited to investigate the distribution of all (and thus also
of dark) matter in the Universe. Lensing results have now become
available over a wide range of scales, from the search for MACHOs in
the Galactic halo, to the mass distribution in galaxies and clusters
of galaxies, and the statistical properties of the large-scale matter
distribution in the Universe. Here, after introducing the concepts of
strong and weak lensing, several applications are outlined, from
strong lensing by galaxies, to strong and weak lensing
by clusters and the lensing properties of the large-scale structure.
\end{abstract}

\firstsection 
\section{Introduction}
Light rays are deflected in gravitational fields, just like massive
particles are. Hence, the deflection of light probes the gravitational
field, and therefore the matter distribution that causes it. Since the
field is independent of the state and nature of the matter generating
it, it provides an ideal tool for studying the total (that is,
luminous and dark) matter in cosmic objects. As we shall see,
gravitational light deflection is used to study cosmic mass
distributions on scales ranging from stars to galaxies, and from clusters
of galaxies to the large-scale matter distribution in the Universe. In
this contribution, I will concentrate on those aspects which are of
particular relevance for learning about the dark matter distribution
in the Universe.

Gravitational lensing describes phenomena of gravitational light
deflection in the weak-field, small deflection limit; strong-field
light deflection (important for light propagation near black holes or
neutron stars) are not covered by gravitational lens (hereafter GL)
theory. The basic theory of gravity, and of light
propagation in a gravitational field is General Relativity, which says
that photons travel along null geodesics of the spacetime metric
(these are described by a second-order differential equation). In GL
theory, several simplifications apply, owing to restriction to weak
fields, and thus small deflections. We shall see the convenience of
those further below.

Gravitational lensing as a whole, and several particular aspects of
it, has been reviewed previously. Two extensive monographs (Schneider,
Ehlers \& Falco 1992, hereafter SEF; Petters, Levine \& Wambsganss
2001, hereafter PLW) describe lensing in all depth, in particular
providing a derivation of the gravitational lensing equations from
General Relativity. Fort \& Mellier (1994) describe the giant luminous
arcs and arclets in clusters of galaxies (see Sect.\ \ref{sc:4.3}),
Paczy\'nski (1996) and Roulet \& Mollerach (1997) review the effects
of gravitational microlensing in the Local Group, whereas the reviews
by Narayan \& Bartelmann (1999) and Wambsganss (1998) provide a
concise and didactical account of GL theory and observations. Much of
this contribution will be focused on weak gravitational lensing, which
has been reviewed recently by Mellier (1999), Bartelmann \& Schneider
(2001), Wittman (2002), van Waerbeke \& Mellier (2003) and Refregier
(2003).

\section{Basics of gravitational lensing}
\subsection{Very brief history of lensing}
The investigation of gravitational light deflection dates back more than
200 years to Mitchel, Cavendish, Laplace and Soldner (see SEF, PLW for
references and much more detail). At that time a metric theory of
gravity was not known, and light was treated as massive particles
moving with the velocity of light. General Relativity, finalized in
1915, predicts a deflection angle twice as large as `Newtonian'
theory, and was verified in 1919 by measuring the deflection of light
near the Solar limb during an eclipse. Soon after, the `lens effect'
was discussed by Lodge, Eddington and Chwolson, i.e.\ the possibility
that light deflection leads to multiple images of sources behind mass
concentrations, or even yields a ring-like image. Einstein, in 1936,
considered in detail the lensing of a source by a star (or a point-mass
lens), and concluded that the angular separation between the two
images would be far too small (of order milliarcseconds) to be
resolvable, so that ``there is no great chance of observing this
phenomenon''. In 1937, Zwicky, instead of looking at lensing by stars in our
Galaxy, considered ``extragalactic nebulae'' (nowadays called
galaxies) as lenses. He noted that they produce angular separations
than can be separated with telescopes. Observing such an effect, he
noted, would furnish an additional test of GR, would allow one to see
galaxies at larger distances (due to the magnification effect), and
to determine the masses of these nebulae acting as lenses. He
furthermore considered the probability of such lens effects and
concluded that about 1 out of 400 distant sources should be affected
by lensing, and therefore predicted that ``the probability that
nebulae which act as gravitational lenses will be found becomes
practically a certainty''. His visions were right on (nearly) all
accounts.

In the mid-1960's, Klimov, Liebes, and Refsdal independently
formulated the basic theory of gravitational lensing, and focused on
astrophysical applications, like determination of masses and
cosmological parameters. For example, Refsdal noticed that the light travel
time along the two rays corresponding to two images of a source, is
proportional to the size of the Universe, thus to $H_0^{-1}$, and that
a measurement of the time delay, possible if the source varies
intrinsically, would allow the determination of the Hubble constant.

\begin{figure}
\centerline{\psfig{figure=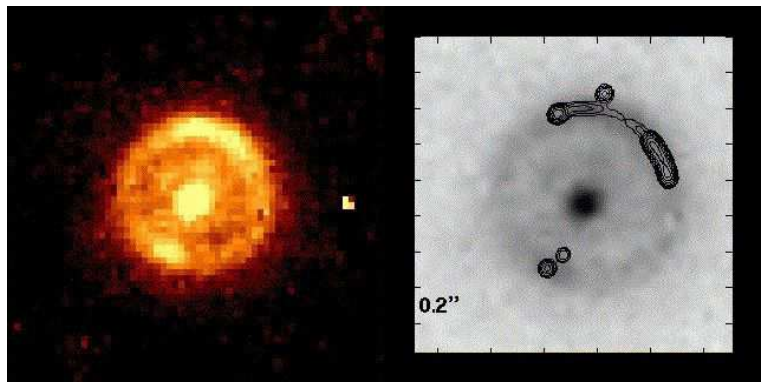,width=10cm}}
\caption{The radio source JVAS B1938+666 shows two radio sources
(contours in the right panel), one of which is mapped into four
components, the other shows a double image; furthermore, the outer
radio contours merge into an arc around the lensing galaxy. The
underlying grey-scale figure, and the left panel, shows a near-IR
iamge of the field, revealing the lens galaxy, as well as the
Einstein-ring image of the galaxy hosting the radio-AGN (from King et
al.\ 1998)} 
\flabel{King}
\end{figure}

In 1979, the first GL system was discovered by Walsh, Carswell \&
Weymann: The two images of the QSO 0957+561 are separated by about 6 
arcseconds, having identical colors, redshifts ($z_{\rm s}=1.41$) and
spectra; both images are radio sources with a core-jet structure on
milli-arcsec scales. Soon thereafter, a galaxy situated between the
two quasar images was detected, with redshift $z_{\rm d}=0.36$, being
member of a cluster. 1980 marks the discovery of the first GL system
with four QSO images, QSO 1115+080, two of which are very closely
spaced. 1986 saw the discovery of a new lensing phenomenon, which had
been predicted long before: the detection of a radio ring, in which
an extended radio source is mapped into a complete ring by an
intervening galaxy (see also Fig.\ \ref{fig:King}). Such {\em Einstein
rings} turn out to yield the most accuracte mass determinations in
extragalactic astronomy.  At present, some 72 multiple-image systems
are known where the major lens component is a galaxy, including about
46 doubles, 20 four-image systems, but also one 3-image, one 5-image,
and one 6-image system.  The first of these were discovered
serendipitously, but since the 1990's, large systematic searches for
such systems were successfully conducted in the optical and, in
particular, radio wavebands.

\begin{figure}
\psfig{figure=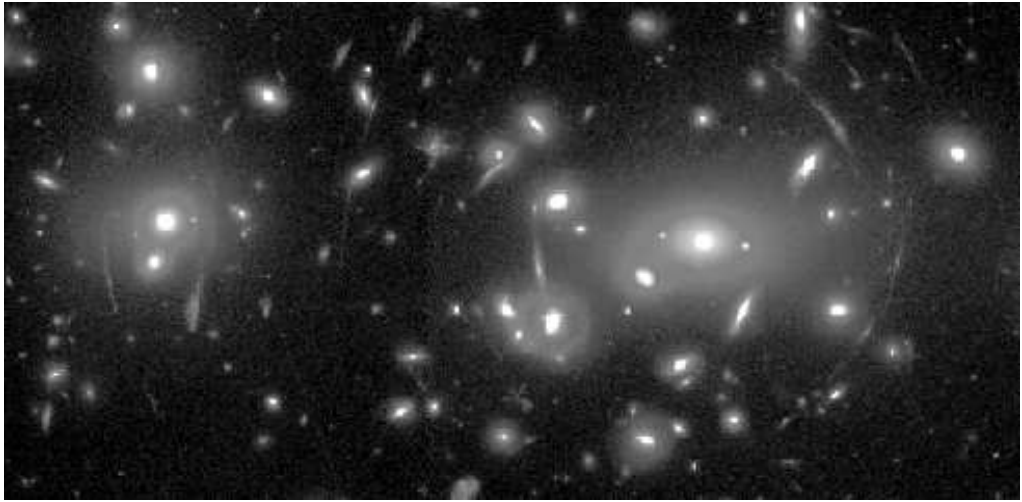,width=0.999\hsize}
\caption{The redshift $z=0.18$ cluster Abell 2218 displays an
enormously rich structure of arcs, highly stretched images of
background galaxies which curve around the main cluster center (seen
to the right of the image center), but also around the secondary mass peak
near the left edge of this WFPC2@HST image. Together with the
identification of several multiply-imaged background sources, these
lensing phenomena have yielded a very detailed mass map of the inner
part of this cluster (Courtesy J.-P. Kneib)
}
\flabel{Fig2}
\end{figure}

In 1986, a new lensing phenomenon was discovered by two independent
groups: strongly elongated, curved features around two clusters of
galaxies. Their extreme length-to-width ratios made them difficult to
interpret; the measurement of the redshift of one of them placed the
source of the arc at a distance well behind the corresponding
cluster. Hence, these giant luminous arcs are images of background
galaxies, highly distorted by the tidal gravitational field of the
cluster. By now, many clusters with giant arcs are known and
investigated in detail, for which in particular the high-resolution of
the HST was essential. Less extremely distorted images of background
galaxies have been named {\em arclets} and can be identified in many
clusters.

If some sources are so highly distorted as those seen in Fig.\
\ref{fig:Fig2}, one expects to see many more sources which are distorted
to a much smaller degree -- such that they can not be identified
individually as lensed images, but that nearby images are distorted in
a similar way, so that the distortion can be identified
statistically. This forms the basis of weak lensing; coherent image
distortions around massive clusters were detected in the early
1990's by Tyson and his group. As shown by Kaiser \& Squires, 
there image distortions can be used to obtain a parameter-free mass
map of clusters. Further weak 
lensing phenomena, such as galaxy-galaxy lensing, have been detected
in the last decade; the weak lensing effect by the large-scale matter
distribution in the Universe, the so-called cosmic shear, was
discovered by several independent teams simultaneously in 2000, and
this has opened up a new window in observational cosmology.

Last but not least, gravitational microlensing in the local group has
been suggested in 1986 by Paczynski as a test of whether the dark
matter in the halo of our Galaxy is made up of compact objects; the
first microlensing events were discovered in 1993 by three different
groups. I refer to the lectures of Prof.\ Sadoulet for a discussion of
microlensing and the results concerning the dark matter in our Milky
Way.

\subsection{Deflection angle and lens equation}
We shall provide here the basic lensing relations; the reader is
encouraged to refer to one of the reviews or books listed in the
introduction for a full derivation of these relations.

\subsubsection{Deflection by a `point mass' $M$}
Consider the deflection of a light ray by the exterior of a
spherically symmetric mass $M$; from the Schwarzschild metric one
finds that a ray with impact parameter $\xi$ 
is deflected by an angle 
\be \ebox{
  \hat\alpha = \frac{4GM}{c^2\,\xi}={2 R_{\rm s}\over \xi} }\; :
\;\; \hbox{Einstein deflection angle;} \;\;
\hbox{$R_{\rm s}$: Schwarzschild radius}
\ee
valid for $R_{\rm s}/\xi\ll 1$, or $\hat\alpha\ll 1$; note that this also
implies that $\Phi/c^2\ll 1$, where $\Phi$ is the
Newtonian gravitational potential. The value for $\hat\alpha$ is twice
the `Newtonian' value derived by Soldner and others and was verified
during the Solar Eclipse 1919! 

\subsubsection{Deflection by a mass distribution}
Since the field equations of General Relativity can be linearized if
the gravitational field is weak, the deflection angle caused by an
extended mass distribution can be calculated as the (vectorial) sum of
the deflections due to its individual mass elements. If the deflection
angle is small (which is implied by the weak-field assumption), the
light ray near the mass distribution will deviate only slightly from the
straight, undeflected ray. In this (`Born') approximation, valid if
the extent of the mass distribution is much smaller than the distances
between source, lens, and observer (the `geometrically thin lens'),
the deflection angle depends solely on the {\em surface mass density}
$\Sigma(\vc\xi)$, defined in terms of the volume density $\rho(\vec
r)$ as 
\begin{equation}
  \Sigma(\vc\xi) \equiv \int\d r_3\,\rho(\xi_1,\xi_2,r_3)\;,
\elabel{2.2}
\end{equation}
with the $r_3$-direction along the line-of-sight. Superposing the
deflections by the mass elements of the lens, one obtains the
deflection angle
\begin{equation} \ebox{
  \hat{\vc\alpha}(\vc\xi) = \frac{4G}{c^2}\,
  \int\d^2\xi'\,\Sigma(\vc\xi')\,
  \frac{\vc\xi-\vc\xi'}{|\vc\xi-\vc\xi'|^2} } \;.
\elabel{2.3}
\end{equation}
The geometrically-thin condition is satisfied in virtually all
astrophysically relevant situations (i.e.~lensing by galaxies and
clusters of galaxies), unless the deflecting mass extends all the way
from the source to the observer, as in the case of lensing by the
large-scale structure. The relevant deflections are small, e.g.,
$\hat\alpha\lesssim 1''$ for galaxies, $\hat\alpha\lesssim 30''$ for
clusters.

\def\Real{{\rm I\mathchoice{\kern-0.70mm}{\kern-0.70mm}{\kern-0.65mm}%
  {\kern-0.50mm}R}}

\subsubsection{The lens equation}
\begin{figure}
\centerline{\psfig{figure=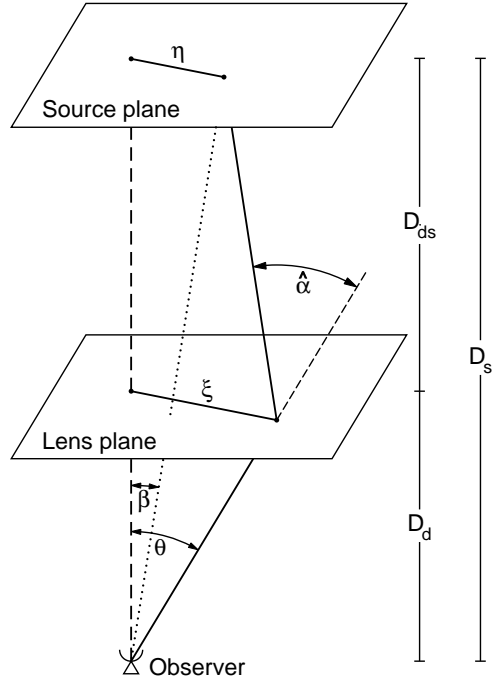,width=0.5\hsize}}
\caption{Sketch of a typical gravitational lens system.}
\flabel{lensgeo}
\end{figure}

The lens equation relates the true position of the source to its
observed position; we define the lens and source plane as planes
perpendicular to the line-of-sight to the deflector, at the distance
$D_{\rm d}$ and $D_{\rm s}$ of the lens and the source, respectively
(see Fig.\ \ref{fig:lensgeo}).  Furthermore, we define the `optical axis'
as a `straight' line through the lens center (the exact definition
does not matter; any change of it represents just an unobservable
translation in the source plane), and its intersections with the lens
and source planes as their respective origins.  Denoting $\vc\xi$ as the
two-dimensional position of the light ray in the lens plane and
$\vc\eta$ as the position of the source (see Fig.\ \ref{fig:lensgeo}),
then from geometry,
\begin{equation}
  \vc\eta = {D_{\rm s} \over D_{\rm d}} \,
  \vc\xi - D_{\rm ds}\hat{\vc\alpha}(\vc\xi)\;.
\elabel{2.4}
\end{equation}
Note that the distances $D$ occuring here are the angular-diameter
distances, since they relate physical transverse separations to
angles. If $\vc\theta$ denotes the angle of a light ray relative to
the optical axis, $\vc\beta$ as the angular position of the unlensed
source (see Fig.\ \ref{fig:lensgeo}),
\be
\vc\eta=D_{\rm s}\vc\beta\; ; \;
\vc\xi=D_{\rm d}\vc\theta\;,
\ee
then
\begin{equation} \ebox{
  \vc\beta = \vc\theta-\frac{D_{\rm ds}}{D_{\rm s}}\,
  \hat{\vc\alpha}(D_{\rm d}\vc\theta) \equiv
  \vc\theta-\vc\alpha(\vc\theta) } \;,
\elabel{3.6}
\end{equation}
where 
$\vc\alpha(\vc\theta)$ is the {\em scaled deflection angle}, which in
terms of the {\em dimensionless surface mass density}
\begin{equation}
\ebox{  \kappa(\vc\theta) := 
  \frac{\Sigma(D_{\rm d}\vc\theta)}{\Sigma_{\rm cr}} }
  \quad\hbox{with}\quad
\ebox{  \Sigma_{\rm cr} = \frac{c^2}{4\pi G}\,
  \frac{D_{\rm s}}{D_{\rm d}\,D_{\rm ds}} } \;,
\elabel{3.7}
\end{equation}
reads as
\begin{equation} \ebox{
  \vc\alpha(\vc\theta) = \frac{1}{\pi}\,
  \int_{\Real^2}\d^2\theta'\,\kappa(\vc\theta')\,
  \frac{\vc\theta-\vc\theta'}{|\vc\theta-\vc\theta'|^2} } \;.
\elabel{3.8}
\end{equation}
Note that the {\em critical surface mass density} $\Sigma_{\rm cr}$
depends only on the distances. Lenses with $\kappa \sim 1$ at some
points are called {\em strong lenses}, and those with $\kappa\ll 1$
everywhere are {\em weak lenses}.  The lens equation
$\vc\beta=\vc\theta-\vc\alpha(\vc\theta)$ is a mapping $\vc\theta
\to\vc\beta$ from the lens plane to the source plane; but in general,
this mapping is non-invertable: for a given source position $\vc\beta$,
the lens equation can have multiple solutions $\vc\theta$ which
correspond to multiple images of a source at $\vc\beta$.

\subsubsection{Deflection and Fermat potentials}
Since $\nabla \ln |\vc\theta|=\vc\theta/|\vc\theta|^2$, the deflection
angle can be written as
\be
\ebox{ \vc \alpha =\nabla\psi} \;,
\quad{\rm with}\quad
\ebox{\psi(\vc\theta)={1\over
\pi}\int_{\Real^2}\d^2\theta'\;\kappa(\vc\theta')\,
\ln|\vc\theta-\vc\theta'|
}\;;
\elabel{def-pot}
\ee
being the {\em deflection potential}; hence, the lens equation
describes a gradient mapping. From
$\nabla^2\ln|\vc\theta|=2\pi\delta_{\rm D}(\vc\theta)$, where
$\delta_{\rm D}$ denotes Dirac's delta-`function',
one finds the
2-D Poisson equation
\be
\ebox{\nabla^2\psi =2\kappa } \;.
\elabel{Poisson}
\ee
Defining the {\it Fermat potential}
\be
\phi(\vc\theta;\vc\beta) ={1\over
2}\rund{\vc\theta-\vc\beta}^2-\psi(\vc\theta)\; , 
\elabel{Fermat-pot}
\ee
where $\vc\beta$ enters as a parameter, one sees that the lens
equation can be written as
\be
\nabla \phi(\vc\theta;\vc\beta)=\vc 0 \;.
\elabel{Fermat}
\ee
Solutions of (\ref{eq:Fermat}) can then be classified, according to
whether the potential $\phi$ has a minimum, maximum, or saddle point
at the solution point $\vc\theta$.

\subsection{Effects of lensing}
\subsubsection{Multiple images}
Multiple images correspond to multiple solutions $\vc\theta$ of the
lens equation for fixed source position $\vc\beta$. For the case of a
point-mass lens,
\be
\hat{\vc\alpha}(\vc\xi)={4GM\over c^2}\,{\vc\xi\over |\vc\xi|^2}\;,
\quad
\vc\beta=\vc\theta-{4 G M D_{\rm ds}\over c^2\,D_{\rm d} D_{\rm s}}
{\vc\theta \over |\vc\theta|^2} =\vc\theta -\theta_{\rm E}^2
{\vc\theta \over |\vc\theta|^2} \;,
\elabel{PMLens}
\ee
where $\theta_{\rm E}$ is the {\em Einstein angle} $\propto
\sqrt{M}$. Note that (\ref{eq:PMLens}) can also be obtained from
(\ref{eq:2.3}) and (\ref{eq:3.6}) by setting
$\Sigma(\vc\xi)=M\,\delta_{\rm D}(\vc\xi)$. 
The lens equation has two solutions, one on either side of
the lens (just solve the quadratic equation for $\theta$), with 
image separation $\Delta\theta\gtrsim 2\theta_{\rm E} \propto
\sqrt{M}$\footnote{Mathematically, substantially larger 
separations can occur if
$|\vc\beta|\gg \theta_{\rm E}$, but this case is astronomically
irrelevant, as explained shortly.}. Hence, the image separation
yields an estimate for the mass of the lens. In general, however, more
complicated mass models are needed to fit the observed image positions
in a gravitational lens system, i.e., to find a mass model and a
source position 
such that $\vc\beta=\vc\theta_i-\vc\alpha(\vc\theta_i)$ is satisfied for
all images $\vc\theta_i$.

\subsubsection{Magnification}
Gravitational light deflection conserves surface brightness; this
follows from Liouville's theorem, noting that light deflection is not
associated with emission or absorption processes. Therefore, $
I(\vc\theta) = I^{(\rm s)}[\vc\beta(\vc\theta)]$, where
$I(\vc\theta)$ and $I^{(\rm s)}(\vc\beta)$ denote the surface brightness
in the image and source plane. Differential light bending causes light
bundles to get distorted; for very small light bundles, the distortion
is described by the Jacobian matrix
\begin{equation}
  {\mathcal A}(\vc\theta) =
  \frac{\partial\vc\beta}{\partial\vc\theta} =
  \left(\delta_{ij} -
    \frac{\partial^2\psi(\vc\theta)}{\partial\theta_i\partial\theta_j}
  \right) = \left(
    \begin{array}{cc}
      1-\kappa-\gamma_1 & -\gamma_2 \\ 
      -\gamma_2 & 1-\kappa+\gamma_1 \\
    \end{array}
  \right)\;,
\elabel{3.11}
\end{equation}
where 
\begin{equation}
  \gamma_1 = \frac{1}{2}(\psi_{,11}-\psi_{,22})\;,\quad
  \gamma_2 = \psi_{,12}\;
\elabel{3.12}
\end{equation}
are the two Cartesian components of the shear (or the tidal
gravitational force). For a small source centered on 
$\vc\beta_0=\vc\theta_0-\vc\alpha(\vc\theta_0)$:
\begin{equation}
  I(\vc\theta) = I^{(\rm s)}\left[
    \vc\beta_0+{\mathcal A}(\vc\theta_0)
    \cdot(\vc\theta-\vc\theta_0)
  \right]\;.
\elabel{3.13}
\end{equation}
Hence, the image of a small circular source with radius $r$ is an
ellipse with semi-axes $\lambda_{1,2}\, r$ where $\lambda_{1,2}$ are
the eigenvalues of $\A$; the orientation of the ellipse is determined by
the shear components $\gamma_{1,2}$.

The area distortion by differential deflection yields a magnification
(since $I$ is unchanged, and ${\rm flux}=I \times \hbox{solid angle}$), 
\be \ebox{
\mu={S\over S_0}=\frac{1}{\det{\mathcal A}} =
\frac{1}{(1-\kappa)^2-|\gamma|^2} } \;, 
\elabel{3.14} 
\ee 
with
$|\gamma|=\sqrt{\gamma_1^2+\gamma_2^2}$. Since $\A$ is different for
different multiple images, the image fluxes are different; the
observed flux ratios yield the image magnification ratios. In
particular, if the image separation in a point mass lens system is
substantially larger than $2\theta_{\rm E}$, which occurs for
$|\vc\beta|\gg \theta_{\rm E}$, the secondary image is very strongly
demagnified and thus invisible.  Flux ratios can in principle 
be used to constrain
lens models, in addition to the image positions, but the
magnifications are affected by small-scale structure in the mass
distribution (we shall come back to this point below), rendering them 
less useful in mass model determinations. Note that the final
expression in (\ref{eq:3.14}) can have either sign; images with
$\mu>0$ have positive parity, those with $\mu<0$ negative
parity. Positive parity images correspond to extrema of the Fermat
potential $\phi$, negative parity images correspond to saddle points
of $\phi$. In the rest of this article, we will always mean the
absolute value of the magnification when writing $\mu$.

\subsubsection{\llabel{3.2}Shape distortions}
The image shape of extended (resolved) sources is changed by lensing,
since the eigenvalues of $\A$ are different in general; rewriting
\be \ebox{
  {\mathcal A}(\vc\theta) = (1-\kappa)
 \left(
    \begin{array}{cc}
      1-g_1 & -g_2 \\ 
      -g_2 & 1+g_1 \\
    \end{array}
  \right) } \;,\; 
\ebox{ g_i={\gamma_i\over (1-\kappa)} } :\;\hbox{reduced shear}\;,
\elabel{redshearM}
\end{equation}
one sees that the shape distortion is determined by the reduced
shear. This in fact forms the basis of {\em weak lensing}. It should
be noted that giant arcs cannot be described by the linearized lens
equation, as they are too big.

\subsubsection{Time delay}
The light travel time along the various ray paths corresponding to
different images is different in general. This implies that 
variations of the source luminosity will show up as flux variations of
the different images at different times, shifted by a time delay
$\Delta t$,
\be
\Delta t = {D_{\rm d}D_{\rm s}\over c\,D_{\rm ds}}(1+z_{\rm d})
\eck{\phi(\vc\theta^{(1)};\vc\beta)-\phi(\vc\theta^{(2)};\vc\beta)} \;,
\elabel{deltat}
\ee
where $\vc\theta^{(1)}$ and $\vc\theta^{(2)}$ are the two image
positions considered. In fact, $\phi(\vc\theta;\vc\beta)$ is, up to an
affine transformation, the light travel time along a ray from the
source at $\vc\beta$ which crosses the lens plane at $\vc\theta$.
Recalling that $\nabla\phi(\vc\theta;\vc\beta)=0$ was equivalent to the
lens equation, we see from (\ref{eq:deltat}) that this is Fermat's
principle in lensing: the light travel time is stationary at physical
images.  Note that $\Delta t \propto H_0^{-1}$, since all the
distances $D$ are $\propto c/H_0$; 
hence, a measurement
of the time delay can be used to determine the Hubble constant,
provided the lens model is sufficiently well known. We shall return to
this issue below.

\subsubsection{General properties of lenses}
If $\Sigma(\vc\xi)$ is a smooth function, then for (nearly) every
source position $\vc\beta$, the number of images is odd (`odd-number
theorem'; Burke 1981).  If in addition, $\Sigma(\vc\xi)$ is
non-negative, then at least one of the images (corresponding to a
minimum in light travel time) is magnified, $\mu\ge 1$ (`magnification
theorem'; Schneider 1984).  The odd-number theorem is violated
observationally: one (usually) finds doubles and quads. The missing
odd image is expected to be close to the center of the lens, where
$\kappa\gg 1$ presumably, meaning that $\mu\ll 1$; hence, this central
image is highly demagnified, and thus not observable.
Both of these theorems can be generalized even to non-thin lenses
(Seitz \& Schneider 1992).

The closed and smooth curves where $\det\A(\vc\theta)=0$ are called
{\em critical curves}. When they are mapped back into the
source plane using the lens equation, the corresponding curves in the
source plane are called {\em caustics}.
\begin{figure} 
\centerline{\psfig{figure=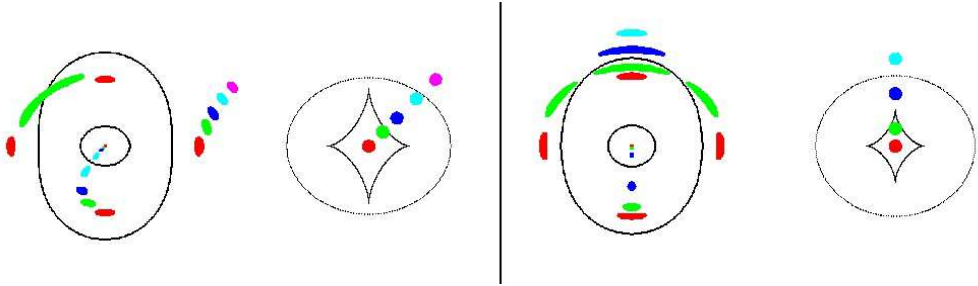,width=0.99\hsize}}
\caption{Illustration of critical curves and caustics for an
elliptical mass distribution, and the image geometry for various
source positions (taken from Narayan \& Bartelmann 1999). In each of
the two panels, the right figure shows the caustic curves, together
with several different source positions; the corresponding image
positions are displayed on the left, together with the critical
curves. One sees that, depending on where the source is located
relative to the caustics, different image multiplicities occur}
\end{figure}
The number of images changes by $\pm 2$ if the source position 
crosses a caustic; then two images appear or merge.  A source close to,
and at the inner side of, a caustic produces two closely separated and
very bright images near the corresponding critical curve.  A source close
to, and on the inner side of a cusp has three bright images close to
the corresponding point on the critical curve. From singularity theory,
one finds that in the limit of very large magnifications, the two
close images on either side of the critical curve have equal
magnification, and thus should appear equally bright. Similarly, of
the three images formed near a cusp, the sum of the magnifications of
the outer two images should equal that of the middle image, with
corresponding consequences for the flux ratios. As we shall see, these
universality relations are strongly violated in observed lens systems,
providing a strong clue for the presence of substructure in the mass
distribution of lens galaxies.

\setcounter{equation}{0}

\section{(Strong) Lensing by galaxies}
The first lensing phenomena detected were multiple images of
distant QSOs caused by the lensing effect of a foreground galaxy. If
the lens is a massive (i.e., $\sim L_*$) galaxy, the corresponding
image separations are $\sim 1''$.  Gravitational lens models can be
used to constrain the mass distribution in (the inner part of) these
lensing galaxies; in particular, the mass inside a circle traced by
the multiple images (or the Einstein ring) can be determined with {\em
very} high precision in some cases. Furthermore, as already mentioned,
time-delays can be used to determine $H_0$.  Mass substructure in
these galaxies can be (and has been) detected, and the interstellar
medium of lens galaxies can be investigated. We shall describe some of
these techniques and results in a bit more detail below.

\subsection{Mass determination}
To obtain accurate mass estimates, one needs detailed models, obtained
by fitting images and galaxy
positions (and fluxes). However, even without these detailed models, 
a simple mass estimate is possible: the mean surface mass density
inside the Einstein radius $\theta_{\rm E}$ of a lens is the critical
surface mass density, so that
\be \ebox{
M(\theta_{\rm E})=\pi (D_{\rm d}\theta_{\rm E})^2\,\Sigma_{\rm cr} }\; .
\elabel{mass}
\ee
An estimate of $\theta_{\rm E}$ is obtained as the radius of the circle
tracing the multiple images (or the ring radius in case of Einstein
ring images). The estimate (\ref{eq:mass}) is exact for axi-symmetric
lenses, and also a very good approximation for less symmetrical ones.

\subsection{Mass models}
The simplest mass model for a galaxy is that of a singular isothermal
sphere (SIS), which is an analytic solution of the Vlasov--Poisson
equation of stellar dynamics (see Binney \& Tremaine 1987) and
whose density profile behaves like $\rho(r)\propto
r^{-2}$, so that the surface mass density is given by
\be
\Sigma(\xi)={\sigma_v^2\over 2G\xi}\; ;\;\;{\rm with}\;\;
\sigma_v:\;\hbox{1-D velocity dispersion.}
\ee
This model is often good enough for rough estimates, in particular
since the inner parts of the radial mass profile of galaxies seem to
closely follow this relation. Multiple images occur for
$\beta<\theta_{\rm E}$, and their separation is
$\Delta\theta=2\theta_{\rm E}$, with
\be
\theta_{\rm E}=4\pi\rund{\sigma_v\over c}^2\rund{D_{\rm
ds}\over D_{\rm s}}
\approx 1\arcsecf 15 \rund{\sigma_v\over 200\,{\rm km/s}}^2\rund{D_{\rm
ds}\over D_{\rm s}}\;.
\ee
Hence, massive ellipticals create image separations of up to $\sim
3''$, whereas for less massive ones, and for spirals, $\Delta\theta$
is of order or below $\sim 1''$. 

However, this simple mass model is unrealistic owing to its diverging
density for $r\to 0$ and its infinite mass; furthermore, it (like all
axi-symmetric models) cannot account for the occurrence of quadruply
imaged sources. More complicated models include some or all of these:
\bi
\item
A finite core size, to remove the central divergence. Applied to
observed systems, the models `like' the core to be very small, in
particular since the third or fifth image is not seen, which needs high
demagnification $\mu\ll 1$ and thus high $\kappa$ near the center.
\item
Elliptical isodensity contours that break the axial symmetry, which 
is needed to explain 4-image systems; those cannot be produced by
symmetric lens models.
\item
External tidal field (shear): A lens galaxy is not isolated, but may
be part of a group or a cluster, and in any case the inhomogeneous
matter distribution between source and lens, and lens and observer
introduces a shear (cosmic shear) of typically a few percent.  This
external influence can be linearized over the region of the galaxy
where multiple images occur, and this yields a uniform shear term in
the lens equation.  
\ei

In fact, any realistic model of a lens consists of at least an
elliptical mass distribution and an external shear. This then yields
the necessary number of free parameters for a lens model: 1 for the
mass scale of the lens (either the Einstein radius, or $\sigma_v$),
2 for the lens position, 2 for the source position, 2 for the lens
ellipticity (axis ratio and orientation), and 2 for the external shear
(a two-component quantity). This can be compared to the number of
observables. In a quad-system, one has $2\times 4$ image positions,
and the 2 coordinates of the lens galaxy. In this case, the number of
observational constraints is larger by one than the number of free
parameters. In addition, one could use the flux ratios of the images
(i.e., the magnification ratios) to constrain the lens model, but as
we shall discuss below, these are not reliable constraints for fitting
a macro-model.

\begin{figure}
\centerline{\psfig{figure=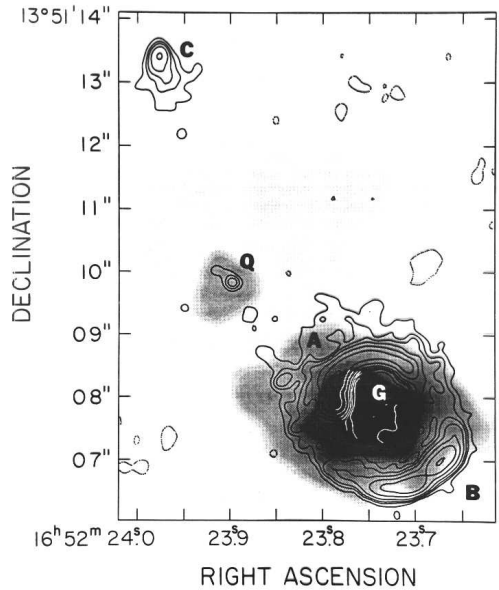,width=7cm}}
\caption{The Southern radio lobe (the Northern one is denoted by C)
of the quasar MG\ 1654+134
($z=1.7$, central component Q coincident with the optical QSO position),  
shown here as contours, 
is mapped into a complete Einstein ring by a foreground galaxy (G),
indicated by the grayscale image. For this lens system, probably the
most datailed mass analysis was performed -- see text 
(image is from Langston et al.\ 1990)}
\flabel{1654}
\end{figure} 

Modeling of strong lens systems which contain Einstein rings yield
much better constrained lens parameters and therefore even more
accurate mass estimates; a beautiful example of this is the radio ring
MG1645+134 (see Kochanek 1995; Wallington et al.\ 1996) around a
foreground elliptical galaxy at $z=0.25$ shown in Fig.\
\ref{fig:1654}.

Results from fitting parametrized mass models to observed multiple
image systems include the following: Many lens systems require quite a
strong external shear, which may be explained by massive
ellipticals being preferentially located in dense regions, i.e.\ in
clusters or groups.  The orientation of the mass ellipticity follows
closely that of light distribution; however, this is not the case for
the magnitude of ellipticity (Keeton et al.\ 1998).  From mass
modelling and detailed spectroscopic studies of lens galaxies, the
latter in combination with stellar dynamical arguments, one finds that
inside the Einstein radius, about half the mass is dark, and half is
baryonic (Treu \& Koopmans 2002; Koopmans \& Treu 2003); therefore, in
massive (lens) galaxies, baryons have strongly affected the mass
profile, owing to their cooling and contraction. These authors have
also shown that over the radius relevant for lensing, the profile is
very well approximated by an isothermal one.

This issue is closely related to the determination of the Hubble
constant from measuring time delays in multiple image systems.  At
present, time delays are known for about 10 lens systems, in some cases
with an accuracy of 1\%. Therefore, $H_0$ can in principle be obtained
by (\ref{eq:deltat}). However, reality turns out to be considerably
more difficult. The major difficulty lies in the so-called mass-sheet
degeneracy, which says that the transformation $\kappa(\vc\theta) \to
\lambda \kappa(\vc\theta)+(1-\lambda)$ of the surface mass density
leaves the image positions and magnification ratios invariant, but
changes the time delay by a factor $\lambda$ (Falco et al.\ 1985).
Essentially, the bracket in (\ref{eq:deltat}) depends on the mean
surface mass density within the annular region around the lens in
which the multiple images are located (Kochanek 2002), and the
mass-sheet degeneracy changes that value. One therefore requires
additional information about the mass profile in galaxies.

As mentioned above, an isothermal profile ($\kappa\propto
\theta^{-1}$) provides a reasonable fit to lens systems. Assuming an
isothermal profile yields values of the Hubble constant of order
$H_0\approx 50\,{\rm km/s/Mpc}$, consistently for the `simple' lens
systems. This is at variance with the value $H_0\approx 72\,{\rm
km/s/Mpc}$ obtained from the Hubble Key Project (Freedman et al.\
2001). On the other hand, the isothermal profile is at best a
reasonable approximation to the real mass profile. Cosmological
simulations yield a cuspy profile, such as the one found by Navarro et
al.\ (1997; herafter NFW). These dark matter profiles are then
modified by baryons cooling inside these halos; the larger the baryon
fraction, the more are the dark matter profiles affected. Kochanek
(2003) pointed out that in order to get a value of $H_0$ from lensing
which is compatible with that from the Hubble Key Project, one would
need a baryon fraction as large as $20\%$ of the total dark
matter in the halo to cool, in order to render the central mass
profiles of lenses steep enough; in effect, that leads to constant
M/L-models within the region where the multiple images are found.
This high fraction of cold baryons in galaxies is at odds
with the local inventories of baryonic mass in galaxies.  At present,
the origin of this discrepancy is not known.

\subsection{Substructure in lens galaxies}
Whereas `simple' lens models can usually fit the image positions {\em
very} well, in most lens systems they are unable to provide a good fit
to the flux ratios. The best known (but by far not worst) case is QSO
1422+231, where several groups have tried, and failed, to obtain a
good lens model explaining image positions and fluxes. Mao \&
Schneider (1998) have provided an analytical argument, based on the
universality of the lens mapping near cusps, 
why one would
not expect to find a smooth model for this system. Since the flux
ratios in this system are most reliably and accurately measured in the
radio, absorption by the ISM in the (elliptical) lens galaxy is
expected to be negligible. We argued that small-scale structure in
the mass distribution can change the magnification, but leave the
image position essentially unchanged; this is due to the fact that the
deflection angle depends on first partial derivatives of the
deflection potential $\psi$, whereas the magnification depends on
$\kappa$ and $\gamma$, and thus on second derivatives of $\psi$; those
are more strongly affected by small-scale structure.

This effect has been known for a long time: the optical and UV
radiation from QSOs comes from a region small enough that even stars
in the lens galaxy can affect their magnification, whereas the
corresponding deflections are of order $10^{-6}$\ arcseconds -- this
{\em microlensing} phenomenon (see, e.g., Wambsganss et al.\ 1990)
shows up as uncorrelated brightness variations in the multiple images,
and has clearly been detected in the QSO 2237+0305 (Schmidt et al.\
2002) and in some of the other lens systems. However, the VLBI images
of QSO 1422+231 are extended, and therefore individual stars cannot
affect their magnification. But massive structures with $M\gtrsim
10^7M_\odot$ can change their magnification. Recall that CDM models of
structure formation actually predict the presence of sub-halos in each
massive galaxy -- the (missing, since unobserved) satellites. As shown
by Dalal \& Kochanek (2002; and references therein), the statistics of
mismatches between observed flux ratios and those predicted by simple
lens models which fit the image positions in 4-image systems is in
agreement with expectations from CDM satellites. Bradac et al.\ (2002)
have explicitly demonstrated that the flux ratio problem in QSO
1422+231 is cured by placing a low-mass halo near one of the QSO
images. Kochanek \& Dalal (2003) have considered, and ruled out,
alternative explanations for the flux ratio mismatches, such as
interstellar scattering. One of the signatures of mass substructure,
first found in investigations of microlensing (Schechter \& Wambsganss
2002), is that the brightest saddle point is expected to be affected
most, in the sense that it has a high probability of being
substantially demagnified. Kochanek \& Dalal have shown that this
particular behavior is seen in a sample of 7 quadruple image lens
systems. This behavior cannot be explained by absorption, scattering
or scintillation by the interstellar medium of the lens galaxy. Hence,
lensing has probably detected the `missing' satellites in galaxy
halos; the observed flux mismatches require a mass fraction in
subclumps of order a few percent of the total lens mass, in accordance
with predictions from CDM simulations.  Bradac et al.\ (2002, 2003)
have generated synthetic lens systems, using model galaxies as
obtained from CDM simulations as deflectors, and have shown that the
resulting image fluxes are at variance with the predictions from
simple lens models, again due to the substructure in the mass
distribution.  In a few of the observed lens systems, a fairly massive
subclump can be identified directly by its luminosity, yielding
further support to this interpretation.

\subsection{Other properties of lens galaxies}
\subsubsection{Evolution}
Early-type galaxies are known to be located on the so-called
fundamental plane (FP), i.e., there is a relation between their
central surface brightness, their effective radius and the velocity
dispersion in these galaxies. The FP has been observed even to high
redshifts, using early-type galaxies in high-redshift clusters; it is
known to evolve with $z$, mainly due to passive evolution of the
stellar population. The lens galaxies form a mass-selected
sample of galaxies not selected for cluster membership, and it is
therefore of great interest 
to see whether they also obey a FP relation. In fact, since lens
galaxies have a well-determined mass scale (or $\sigma_v$, after
fitting an isothermal mass model), and are spread over a large
redshift range, they are ideal for FP research. Rusin et al.\ (2003)
have found from a sample of 28 lenses that the evolution of the FP is
compatible with passive stellar evolution, and that it favours a high
redshift for the epoch of star formation in these galaxies.
 
\subsubsection{The interstellar matter in lens galaxies}
Multiple image systems provide us with views of the same source along
different lines-of-sight. Excluding time-delay effects in connection
with spectral variability, as well as differential magnification,
spectral differences between the images can
then only be caused by propagation effects. In particular, one can
study the properties of the dust in lens galaxies, as color
differences between images can be attributed to different extinction
and reddening along the different lines-of-sight through the 
lens galaxy. Falco et
al.\ (1999) have investigated 23 gravitational lens
galaxies over the range $0\lesssim z_{\rm d}\lesssim 1$. Given that
most lens galaxies are early types, they found a small median
differential extinction of $\Delta E(B-V)\sim 0.05$, with slightly
larger (smaller) values for radio- (optically-)selected lens systems. 
The lack of a clear correlation with the separation of the image from
the lens center points towards patchy extinction. Two spiral lens
galaxies show a substantially larger extinction. The extinction law,
i.e., the relation between extinction and reddening, varies between
different lens galaxies over quite some range; the Galactic extinction
law is therefore by no means universally applicable.

\section{Weak gravitational lensing}
\setcounter{equation}{0}
Multiple images, microlensing (with appreciable magnifications) and
arcs in clusters are phenomena of {\em strong lensing}. In {\em weak
gravitational lensing}, the Jacobi matrix $\A$ is very close to the
unit matrix, which implies weak distortions and small magnifications.
Those cannot be identified in individual sources, but only in a
statistical sense; the basics of these effects will be described in
this section, and several applications will be dicussed in later
sections.

\subsection{Distortion of faint galaxy images}
Images of distant, extended sources are distorted in shape and size;
this is described by the locally
linearized lens equation around the image center $\vc\theta_0$,
\be
\vc \beta-\vc\beta_0={\mathcal A}(\vc\theta_0)
    \cdot(\vc\theta-\vc\theta_0)\;,
\ee
where $\vc\beta_0=\vc\beta(\vc\theta_0)$, with the Jacobian
(\ref{eq:redshearM}), and the invariance of surface brightness
(\ref{eq:3.13}).  Recall that the shape distortion is described by the
(reduced) shear which is a two-component quantity, most conveniently
written as a complex number,
\be
\gamma=\gamma_1 +{\rm i}\gamma_2=|\gamma|\,{\rm e}^{2{\rm i}\vp}\;;
\quad g=g_1+{\rm i} g_2=|g|\,{\rm e}^{2{\rm i}\vp}\;;
\ee
its amplitude describes the degree of distortion, whereas its phase
$\vp$ yields the direction of distortion. The reason for the factor `2' in
the phase is the fact that an ellipse transforms into itself after a
rotation by $180^\circ$.
Consider a circular source with radius $r$; mapped by the local Jacobi
matrix, its image is an ellipse, with semi-axes
\[
{r\over 1-\kappa-|\gamma|}={r\over (1-\kappa)(1-|g|)}
\quad ;\quad{r\over 1-\kappa+|\gamma|}={r\over (1-\kappa)(1+|g|)} \;
\]
and the major axis encloses an angle $\vp$ with the positive
$\theta_1$-axis.  Hence, if circular sources could be identified, the
measured image ellipticities would immediately yield the value of the
reduced shear, through the axis ratio
\[
|g|={1-b/a \over 1+b/a}\quad \Leftrightarrow \quad
{b\over a}={1-|g|\over 1+|g|}
\]
and the orientation of the major axis $\vp$.  However, faint galaxies
are not intrinsically round, so that the observed image ellipticity is
a combination of intrinsic ellipticity and shear.  The strategy to
nevertheless obtain an estimate of the (reduced) shear consists in
locally averaging over many galaxy images, assuming that the intrinsic
ellipticities are {\em randomly oriented}. In order to follow this
strategy, one needs to clarify first how to define `ellipticity' for a
source with arbitrary isophotes (faint galaxies are not simply
elliptical); in addition, seeing caused by atmospheric turbulence will
blur -- and thus circularize -- observed images.  We will consider
these issues in turn.

\subsection{Measurements of shapes and shear}
Let $I(\vc\theta)$ be the brightness distribution of an image, assumed to
be isolated on the sky; the
center of the image can be defined as 
\begin{equation}
  \bar{\vc\theta} \equiv
  \frac{\int\!\d^2\theta\,q_I[I(\vc\theta)]\,\vc\theta}
       {\int\!\d^2\theta\,q_I[I(\vc\theta)]}\;,
\elabel{4.1}
\end{equation}
where $q_I(I)$ is a suitably chosen weight function; e.g., if
$q_I(I)=I\,{\rm H}(I-I_{\rm th})$, $\bar{\vc\theta}$ would be the
center of light within a limiting isophote of the image (where H denotes
the Heaviside step function).
We next define the tensor of second brightness moments,
\begin{equation}
  Q_{ij} = \frac{
    \int\!\d^2\theta\,q_I[I(\vc\theta)]\,
    (\theta_i-\bar\theta_i)\,(\theta_j-\bar\theta_j)
  }{
    \int\!\d^2\theta\,q_I[I(\vc\theta)]
  }\;,\quad i,j\in \{1,2\}\;.
\elabel{4.2}
\end{equation}
Note that 
for an image with circular isophotes, $Q_{11}=Q_{22}$, and $Q_{12}=0$.
The trace of $Q$ describes the size of the image, whereas the
traceless part of $Q_{ij}$ contains the ellipticity information.

From $Q_{ij}$, one defines two complex ellipticities,
\begin{equation} \ebox{
  \chi \equiv \frac{Q_{11}-Q_{22}+2{\rm i}Q_{12}}
  {Q_{11}+Q_{22}} }\; \;\;
{\rm and} \;\; \ebox{
  \epsilon \equiv \frac{Q_{11}-Q_{22}+2{\rm i}Q_{12}}
  {Q_{11}+Q_{22}+2(Q_{11}Q_{22}-Q_{12}^2)^{1/2}}  } \;.
\elabel{4.10}
\end{equation}
Both of them have the same phase (because of the same numerator), but a
different absolute value;
for an image with elliptical isophotes of axis ratio $r\le 1$, one
obtains 
\be
|\chi|={1-r^2\over 1+r^2}\quad ;\quad
|\eps|={1-r\over 1+r}\;.
\ee
Which of these two definitions is more convenient depends on the
context; one can easily transform one into the other,
\be
  \epsilon = \frac{\chi}{1+(1-|\chi|^2)^{1/2}}\;,\quad
  \chi = \frac{2\epsilon}{1+|\epsilon|^2}\;.
\elabel{4.11}
\end{equation}

\subsubsection{From source to image ellipticities}
In total analogy, one defines the second-order brightness tensor $Q^{(\rm
s)}_{ij}$, and the complex ellipticities $\chi^{(\rm s)}$ and
$\epsilon^{(\rm s)}$ for the unlensed source.
From 
\be  
Q^{(\rm s)}_{ij} = \frac{
    \int\!\d^2\beta\,q_I[I^{(\rm s)}(\vc\beta)]\,
    (\beta_i-\bar\beta_i)\,(\beta_j-\bar\beta_j)
  }{
    \int\!\d^2\beta\,q_I[I^{(\rm s)}(\vc\beta)]
  }\;,\quad i,j\in \{1,2\}\;,
\elabel{4.2a}
\end{equation}
one finds with $\d^2\beta=\det\A\,\d^2\theta$,
$\vc\beta-\bar{\vc\beta} =\A\rund{\vc\theta-\bar{\vc\theta}}$ 
that 
\begin{equation}
  Q^{(\rm s)} = \A \,Q\,\A^T = 
  \A \,Q\,\A\;,
\elabel{4.5}
\end{equation}
where $\A\equiv\A(\bar{\vc\theta})$ is the Jacobi
matrix of the lens equation at position $\bar{\vc\theta}$.
Using the definitions of the complex ellipticities, one finds the
transformations: 
\begin{equation} \ebox{
  \chi^{(\rm s)} = \frac{\chi-2g+g^2\chi^*}{1+|g|^2-2\Re(g\chi^*)}  }
\; ;\quad  \ebox{
  \epsilon^{(\rm s)} = \left\{\begin{array}{ll}
    \displaystyle\frac{\epsilon-g}{1-g^*\epsilon} &
    \quad\hbox{if}\; |g|\le1\;; \\
    \\
    \displaystyle\frac{1-g\epsilon^*}{\epsilon^*-g^*} & 
    \quad\hbox{if}\; |g|>1 \;.\\
  \end{array}\right. }
\label{eq:4.6}
\end{equation}
The inverse transformations are
obtained by interchanging source and image
ellipticities, and $g\to -g$ in the foregoing equations.

\subsubsection{Estimating the (reduced) shear}
In the following we make the assumption that the intrinsic orientation
of galaxies is random,
\begin{equation}
  \mathrm{E}\rund{\chi^{(\rm s)}} = 0 = \mathrm{E}\rund{\epsilon^{(\rm s)}}\;,
\elabel{4.13}
\end{equation}
which is expected to be valid since there should be no direction
singled out in the Universe. This then implies that the expectation
value of $\eps$ is [as obtained by averaging the transformation law
(\ref{eq:4.6}) over the phase of the intrinsic source orientation
(Schramm \& Kaiser 1995; Seitz \& Schneider 1997)]
\be \ebox{
{\rm E}(\epsilon) = \left\{\begin{array}{ll}
    \displaystyle g &
    \quad\hbox{if}\; |g|\le1 \\
    \\
    \displaystyle 1/g^* & 
    \quad\hbox{if}\; |g|>1 \;.\\
  \end{array}\right. }
\label{eq:4.6a}
\end{equation}
This is a remarkable result, since it shows that each image
ellipticity provides an  unbiased estimate of the
local shear, though a very noisy one. The
noise is determined by the intrinsic ellipticity dispersion
\[
\sigma_\eps=\sqrt{\ave{\eps^{(\rm s)}\eps^{(\rm s)*}}}\;.
\]
This noise can be beaten down by averaging over many galaxy
images. Fortunately, we live in a Universe where the sky is `full of
faint galaxies', as was impressively demonstrated by the Hubble Deep
Field images (Williams et al.\ 1996). Hence, the accuracy of a shear
estimate depends on the local number density of galaxies for which a
shape can be measured. In order to obtain a high density, one requires
deep imaging observations. As a rough guide, on a 3 hour exposure
under excellent observing conditions with a 3-meter class telescope,
about 30 galaxies per arcmin$^2$ can be used for a shape measurement.

Note that in the weak lensing regime, $\kappa\ll 1$, $|\gamma|\ll 1$,
one finds
\begin{equation}
  \gamma \approx g \approx 
  \langle\epsilon\rangle \approx
  \frac{\langle\chi\rangle}{2}\;.
\label{eq:4.18}
\end{equation}

\subsection{Problems in measuring shear, and their solutions}
\subsubsection{Major problems}
\bi
\item
Seeing, that is the finite size of the point spread function (PSF),
circularizes images; this effect is severe since faint galaxies
(i.e.\ those at a magnitude limit for which the number density is of
order 30 per arcmin$^2$) are not larger than the typical seeing
disk. Therefore, weak lensing requires imaging with very good seeing.
\item
The PSF is not circular, owing to e.g., wind shake of the telescope,
or tracking errors. However, an anisotropic PSF causes round sources
to appear elliptical, and thus mimics shear.
\item
Galaxy images are {\em not} isolated, and therefore the integrals in
the definition of $Q_{ij}$ have to be cut-off at a finite
radius. Hence, one usually uses a weight function $q$ which depends
explicitly on $|\vc \theta-\bar\vc\theta|$; however, this modifies the
transformation (\ref{eq:4.6}) between image and source ellipticity.
\item 
The sky noise, i.e.\ the finite brightness of the night sky,
introduces a noise component in the measurement of image ellipticities
from CCD data, so that only for high-S/N objects can a shape be
measured.
\item
Distortion by telescope and camera optics renders the coaddition of
exposures complex; one needs to employ remapping, using accurate
astrometry, and sub-pixel coaddition.
\ei

Depending on the science application, the shear one wants to measure
is of order a few percent, or even smaller. Essentially all of the
effects mentioned can introduce ellipticities of the same order in the
measured images if they are not carefully taken into account.

\subsubsection{Solutions}
In order to deal with these issues, specific software has been
developed. The one that has been mostly used up to now is the Kaiser
et al.\ (1995, KSB) method, or its implementation
IMCAT. Its basic idea is as follows: the image ellipticity will be
determined by the intrinsic ellipticity, the (reduced) shear, the size
of the PSF and its anisotropy. Note that the PSF can be investigated
by identifying stars (that is: point sources) on the images.  The
response of $\chi$ to the shear depends on the size of the source --
for small sources, blurring by seeing reduces the response to a large
degree. The size of the sources is estimated from the size of the
seeing-convolved images. In addition, the response of $\chi$ to a PSF
anisotropy also depends on the image size. The KSB method, an
essential part of which 
was put forward by Luppino \& Kaiser (1997; for
a complete derivation, see Sect.\ 4.6.2 of Bartelmann \& Schneider
2001), results in the relation 
\be\ebox{
   \chi^{\rm obs}_\alpha= \hat{\chi}^0_\alpha + 
   P^{\rm sm}_{\alpha\beta}
  q_\beta+P^g_{\alpha\beta}g_\beta } \;,
\label{eq:4.85}
\end{equation}
where $\hat{\chi}^0$ is the ellipticity of the source convolved with
the isotropic part of the PSF and therefore, its expectation value
vanishes, according to the assumption of randomly oriented sources,
${\rm E}(\hat{\chi}^0)=0$. The tensor $P^{\rm sm}_{\alpha\beta}$
describes the response of the image ellipticity to the PSF anisotropy,
which is quantified by the (complex) ellipticity $q_\beta$ of the PSF,
as measured from stars.  $P^g_{\alpha\beta}$ is a tensor which
describes the response of the image ellipticity to an applied shear.
Both, $P^{\rm sm}_{\alpha\beta}$ and $P^g_{\alpha\beta}$ are
calculated for each image individually; they depend on higher-order
moments of the image brightness distribution and the size of the PSF.
Detailed simulations (e.g., Erben et al.\ 2001; Bacon et al.\ 2001)
have shown that the KSB method can measure shear with better than $\sim
10\%$ accuracy. Several other methods for measuring image shapes in
order to obtain an estimate of the local shear have been developed and
some of them have already been applied to observational data (Bonnet
\& Mellier 1995; Kuijken 1999; Kaiser 2000; Bernstein \& Jarvis 2002;
Refregier \& Bacon 2003).

\subsection{Magnification effects}
The magnification caused by the differential light bending
changes the apparent brightness of sources; this leads to two
effects:
\ben
\item
The observed flux $S$ from a source is changed from its unlensed value
$S_0$ according to $S=\mu\,S_0$; if $\mu >1$, sources appear brighter
than they would without an intervening lens.
\item
A population of sources in the unlensed solid angle $\omega_0$ is spread
over the solid angle $ \omega=\mu\omega_0$ due to the magnification.
\een
These two effects affect the number counts of sources differently;
which one of them wins depends on the slope of the number counts; one
finds 
\begin{equation} \ebox{
  n(>S,\vc\theta,z) = \frac{1}{\mu(\vc\theta,z)}\,
  n_0\left(>\frac{S}{\mu(\vc\theta,z)},z\right) } \;,
\elabel{4.38}
\end{equation}
where $n(>S,z)$ and $n_0(>S,z)$ are the lensed and unlensed cumulative
number counts of sources, respectively. The first argument of $n_0$
accounts for the change of the flux, whereas the prefactor in
(\ref{eq:4.38}) stems from the change of apparent solid angle.

As an illustrative example, we consider the case that the  source
counts follow a power law, 
\begin{equation}
  n_0(>S) = a\,S^{-\alpha}\;;
\elabel{4.41}
\end{equation}
one then finds for the lensed counts in a region of the sky with
magnification $\mu$:
\be  
\frac{n(>S)}{n_0(>S)} = \mu^{\alpha-1}\;,
\elabel{4.43}
\end{equation}
and therefore, if $\alpha>1$ ($<1$), source counts are enhanced
(depleted); the steeper the counts, the stronger the effect. In the
case of weak lensing, where $|\mu-1|\ll 1$, one probes the source
counts only over a small range in flux, so that they can always be
approximated (locally) by a power law.

One important example is provided by the lensing of QSOs. The QSO
number counts are steep at the bright end, and flat for fainter
sources. This implies that in regions of magnification $>1$, bright
QSO should be overdense, faint ones underdense.
This {\em magnification bias} is the reason why the fraction of lensed
sources is {\em much} higher in bright QSO samples than in fainter ones!

\subsection{Tangential and cross component of shear}
\subsubsection{The shear components}
The shear components $\gamma_1$ and $\gamma_2$ are defined relative to
a reference Cartesian coordinate frame. Note that the shear is {\em
not} a vector, owing to its transformation properties under rotations,
which is the same as that of the linear polarization; it is therefore
called a {\em polar}. In analogy with vectors, it is often useful to
consider the shear components in a rotated reference frame, that is,
to measure them w.r.t.\ a different direction; for example, the arcs
in clusters are tangentially aligned, and so their ellipticity is
oriented approximately tangent to the radius vector in the cluster.

If $\phi$ specifies a direction, one defines the {\em tangential} and
{\em cross components} of the shear {\em relative to this direction} as
\be \ebox{
\gamma_{\rm t}=-\Re\eck{\gamma\,{\rm e}^{-2{\rm i}\phi}} \quad,
\quad
\gamma_\times=-\Im\eck{\gamma\,{\rm e}^{-2{\rm i}\phi}} } \;.
\ee
For example, in the case of a circularly-symmetric matter distribution,
the shear at any point will be oriented tangent to the direction
towards the center of symmetry.  Thus in this case, choose $\phi$ to be
the polar angle of a point; then, $\gamma_\times=0$. In full analogy
to the shear, one defines the tangential and cross components of an
image ellipticity, $\eps_{\rm t}$ and $\eps_\times$.

\subsubsection{Minimum lens strength for its weak lensing detection}
As a first application of this decomposition, we consider how massive
a lens needs to be in order to produce a detectable weak lensing
signal. For this purpose, consider a lens modeled as an SIS with
one-dimensional velocity dispersion $\sigma_v$.  In the annulus
$\theta_{\rm in}\le\theta\le\theta_{\rm out}$, centered on the lens,
let there be $N=n\pi(\theta_{\rm out}^2-\theta_{\rm in}^2) $ galaxy
images with position $\vc\theta_i=\theta_i(\cos\phi_i,\sin\phi_i)$ and
(complex) ellipticities $\eps_i$.  For each one of them, consider the
tangential ellipticity
\be
\eps_{{\rm t}i}=-\Re\rund{\eps_i\,{\rm e}^{-2 {\rm i}\phi_i}}\;.
\ee
Next we define a statistical quantity to measure the degree of
tangential alignment of the galaxy images, and thus the lens strength:
\begin{equation}
  X \equiv \sum_{i=1}^N a_i\,\epsilon_{{\rm t}i}\;,
\label{eq:4.52}
\end{equation}
where the factors $a_i=a(\theta_i)$ are arbitrary at this point, and
will later be chosen such as to maximize the signal-to-noise ratio of
this estimator. The shear of an SIS is given by
\be
\gamma_{\rm t}(\theta)={\theta_{\rm E}\over 2\theta}=
2\pi\rund{\sigma_v\over c}^2\rund{D_{\rm
ds}\over D_{\rm s}}\,{1\over \theta}\;.
\ee
The expectation value of the ellipticity is ${\rm
E}(\epsilon_{{\rm t}i})=\gamma_{\rm t}(\theta_i)$, so that
$
{\rm E}(X)=\theta_{\rm E}\sum_ia_i/(2\theta_i)$.
Since $\epsilon_{{\rm t}i}=\epsilon_{{\rm t}i}^{(\rm s)}+
\gamma_{{\rm t}}(\theta_i)$ in the weak lensing regime, one has 
${\rm E}(\epsilon_{{\rm t}i}\epsilon_{{\rm t}j})
= \gamma_{{\rm t}}(\theta_i)\gamma_{{\rm t}}(\theta_j)
+\delta_{ij}\sigma_\epsilon^2/2$, thus
\begin{equation}
  {\rm E}(X^2) = \sum_{i,j=1}^Na_ia_j\,
  {\rm E}(\epsilon_{{\rm t}i}\epsilon_{{\rm t}j}) =
  [{\rm E}(X)]^2+\frac{\sigma_\epsilon^2}{2}\sum_{i=1}^Na_i^2\;.
\elabel{4.53}
\end{equation}
Therefore, the
signal-to-noise ratio for a detection of the lens is
\begin{equation}
  \frac{{\rm S}}{{\rm N}} = 
  \frac{\theta_{\rm E}}{\sqrt{2}\sigma_\epsilon}\,
  \frac{\sum_ia_i\,\theta_i^{-1}}{\sqrt{\sum_ia_i^2}}\;.
\elabel{4.54}
\end{equation}
The $a_i$ can now be chosen so as to maximize S/N; from differentiation
of S/N, one finds a maximum if $a_i\propto 1/\theta_i$.
Then, performing the ensemble average over the galaxy positions, one
finally obtains:
\begin{eqnarray}
 \frac{{\rm S}}{{\rm N}} &=&
 \frac{\theta_{\rm E}}{\sigma_\epsilon}\,\sqrt{\pi n}\,
 \sqrt{\ln(\theta_{\rm out}/\theta_{\rm in})}
 \label{eq:4.55} \\ &=&
 8.4\,
 \left(\frac{n}{30\,{\rm arcmin}^{-2}}\right)^{1/2}
 \left(\frac{\sigma_\epsilon}{0.3}\right)^{-1}
 \left(\frac{\sigma_v}{600\,{\rm km\,s}^{-1}}\right)^2
 \nonumber\\ &\times& \phantom{8.4\,}
 \left(
   \frac{\ln(\theta_{\rm out}/\theta_{\rm in})}{\ln10}
 \right)^{1/2} 
 \left\langle\frac{D_{\rm ds}}{D_{\rm s}}\right\rangle\;.
 \nonumber 
\end{eqnarray}
From this consideration we conclude that clusters of galaxies with
$\sigma_v\gtrsim 600\,{\rm km/s}$ can be detected
with sufficiently large S/N by weak lensing, but
individual galaxies ($\sigma_v\sim 200\,{\rm km/s}$) are too weak as
lenses to be detected individually.

\subsection{Galaxy-galaxy lensing}
Whereas galaxies are not massive enough to show a weak lensing signal
individually, the signal of many galaxies can be (statistically)
superposed.  Consider sets of foreground (lens) and background
galaxies; on average, in a foreground-background galaxy pair, the
ellipticity of the background galaxy will be oriented preferentially
in the direction tangent to the connecting line. In other words, if
$\vp$ is the angle between the major axis of the background galaxy and
the connecting line between foreground and background galaxy, values
$\pi/4\le \vp\le \pi/2$ should be slighly more frequent than
$0\le\vp\le\pi/4$.  The mean tangential ellipticity $\ave{\eps_{{\rm
t}} (\theta)}$ of background galaxies relative to the direction
towards foreground galaxies measures the mean tangential shear at this
separation. 

The strength of this mean tangential shear measures mass properties of
the galaxy population selected as potential lenses. In order to
properly interpret the lensing signal, one needs to know the redshift
distribution of the foreground and background galaxies.  Furthermore,
one needs to assume a relation between the lens galaxies' luminosity
and mass properties (such as a Faber--Jackson type of relation),
unless the sample is so large that one can finely bin the galaxies
with respect to their luminosities; this requires of course redshift
information. In this way, the velocity dispersion $\sigma_*$ of an
$L_*$-galaxy can be determined from galaxy-galaxy
lensing. Furthermore, galaxy-galaxy lensing provides a highly valuable
tool to study the mass distribution of galaxy halos at distances from
their centers which are much larger than the extent of luminous
tracers, such as stars and gas -- or ask the question of where the
galaxy halos `end'.

Whereas the first detection of galaxy-galaxy lensing (Brainerd et al.\
1996) was based on a single field with $\sim 9'$ sidelength, much larger
surveys have now become available, most noticibly the SDSS (Fischer et
al.\ 2000; McKay et al.\ 2001). These large data sets have allowed the
splitting of the lens galaxies into subsamples and to investigate
their properties separately. From this it was verified that early-type
galaxies have a larger mass than spiral galaxies with the same luminosity,
and that this behaviour extends to large radii. Furthermore, the
lensing signal for early-type galaxies can be detected out to much
larger scales than for late-types. The interpretation of this result
is not unique: it either can mean that ellipticals have a more
extended halo than spirals, or that the lens signal from ellipticals,
which tend to be preferentially located inside groups and clusters,
arises in fact from the host halo in which they reside
(see also Guzik \& Seljak 2002). Indeed, when the lens galaxy sample
is divided into those living in high- and low-density environments,
the former ones have a significantly more extended lensing signal.

What the galaxy-galaxy signal really measures is the relation between
light (galaxies) and mass. In its simplest terms, this relation can be
expressed by a bias factor $b$ and the correlation coefficient
$r$. Schneider (1998) and van Waerbeke (1998) pointed out that lensing
can be used to study the bias factor as a function of scale and
redshift, by correlating the lensing signal with the number density of
galaxies. In fact, as shown in Hoekstra et al.\ (2002b), both $b$ and
$r$ can be expressed in terms of the galaxy-galaxy lensing signal, the
angular correlation function of the (lens) galaxies and the cosmic
shear signal (see Sect.\ 6 below). Applying this method to the
combination of the Red-Sequence Cluster Survey and the VIRMOS-DESCART
survey, they derived the scale dependence of $b$ and $r$; on large
(linear) scales, their results are compatible with constant values,
whereas on smaller scales it appears that both of these functions vary.
Future surveys will allow much more detailed studies on the relation
between mass and light, and therefore determine the biasing properties
of galaxies from observations directly. This is of course of great
interest, since the unknown behavior of the biasing yields the 
uncertainty in the transformation of the power spectrum of the galaxy
distribution, as determined from extensive galaxy redshift surveys, to
that of the underlying mass distribution. Hence, weak lensing is able
to provide this crucial calibration.

\setcounter{equation}{0}
\section{Lensing by clusters of galaxies}
\subsection{Introduction}
Clusters are the most massive bound structures in the Universe; this,
together with the (related) fact that their dynamical time scale
(e.g., the crossing time) is not much smaller than the Hubble time -- so
that they retain a `memory' of their formation -- render them of
particular interest for cosmologists. The evolution of their
abundance, i.e., their comoving number density as a function of mass
and redshift, is an important probe for cosmological models.
Furthermore, they form signposts of the dark matter distribution in
the Universe.  Clusters act as laboratories for studying the evolution
of galaxies and baryons in the Universe. In fact, clusters were
(arguably) the first objects for which the presence of dark matter has
been concluded (by Zwicky in 1933).

\subsection{The mass of galaxy clusters}
Cosmologists can predict the abundance of clusters as a function of
their mass (e.g., using numerical simulations); however, the mass of a
cluster is not directly observable, but only its luminosity, or the
temperature of the X-ray emitting intra-cluster medium. Therefore, in
order to compare observed clusters with the cosmological predictions,
one needs a way to determine their masses.
Three principal methods for determining the mass of galaxy clusters
are in use:
\bi
\item
Assuming virial equilibrium, the observed velocity distribution
of galaxies in clusters can be converted into a mass estimate; this
method typically requires assumptions about the statistical
distribution of the anisotropy of the galaxy orbits.
\item
The hot intra-cluster gas, as visible through its Bremsstrahlung in
X-rays, traces the gravitational potential of the cluster. Under
certain assumptions (see below), the mass profile can be constructed
from the X-ray emission.
\item
Weak and strong gravitational lensing probe the projected mass
profile of clusters; this will be described further below.
\ei

All three methods are complementary; lensing yields the line-of-sight
projected density of clusters, in contrast to the other two methods
which probes the mass inside spheres.
On the other hand, those rely on equilibrium (and symmetry)
conditions.

\subsubsection{X-ray mass determination of clusters}
The intracluster gas emits via Bremsstrahlung; the emissivity depends
on the gas
density and temperature, and, at lower $T$, on its chemical composition.
Assuming that the gas is in hydrostatic equilibrium in the potential well of
cluster, the gas pressure $P$ must balance gravity, or
\[
\nabla P=-\rho_{\rm g}\,\nabla \Phi \;,
\]
where $\Phi$ is the gravitational potential and $\rho_{\rm g}$ is the
gas density. 
In the case of spherical symmetry, this becomes
\[
{1\over \rho_{\rm g}}\,{\d P\over \d r}=-{\d\Phi\over \d r}
=-{G\,M(r)\over r^2}\;.
\]
From the X-ray brightness profile and temperature measurement, $M(r)$,
the total mass inside $r$ (dark plus luminous) can then be determined,
\be \ebox{
M(r)=-{\kB T r^2\over G\mu m_{\rm p}}
\rund{{\d\ln\rho_{\rm g}\over \d r} 
+ {\d\ln T\over \d r}} } \;.
\elabel{6.37}
\ee
However, the two major X-ray satellites currently operating, Chandra
\& XMM-Newton, have revealed that at least the inner regions of
clusters show a considerably more complicated structure than implied
by hydrostatic equilibrium. In some cases, the intracluster medium is
obviously affected by a central AGN, which produces additional energy
and entropy input. Cold fronts, with very sharp edges, and shocks have been
discovered, most likely showing ongoing merger events. The temperature
and metallicity appear to be strongly varying functions of position.
Therefore, mass estimates of central parts of clusters from X-ray
observations require special care.

\subsection{\llabel{4.3}Luminous arcs \& multiple images}
Strong lensing effects in clusters show up in the form of giant
luminous arcs, strongly distorted arclets, and multiple images of
background galaxies. Since strong lensing occurs only in the central
part of clusters (typically corresponding to $\sim 50 h^{-1}\,{\rm
kpc}$), it can be used to probe only their inner mass 
structure. However, strong lensing yields by far the most accurate
central mass determinations; in some favourable cases with many strong
lensing features (such as for Abell 2218; see Fig.\ \ref{fig:Fig2}),
accuracies better than $\sim 10\%$ can be achieved.

\subsubsection{First go: $M(\le \theta_{\rm E})$}
Giant arcs occur where the distortion (and magnification) is very
large, that is near critical curves.  To a first approximation,
assuming a spherical mass distribution, the location of the arc
relative to
the cluster center (which usually is assumed to coincide with the
brightest cluster galaxy) yields the Einstein radius of the cluster,
so that the mass estimate (\ref{eq:mass}) can be applied.  Therefore,
this simple estimate yields the mass inside the arc radius.  However,
this estimate is not very accurate, perhaps good to within $\sim
50\%$. Its reliability depends on the level of asymmetry and
substructure in the cluster mass distribution (Bartelmann
1995). Furthermore, it is likely to overestimate the mass in the mean,
since arcs preferentially occur along the major axis of clusters. Of
course, the method is very difficult to apply if the center of the cluster
is not readily identified or if it is obviously bimodal.  For these
reasons, this simple method for mass estimates is not regarded as
particularly accurate.

\subsubsection{Detailed modelling}
The mass determination in cluster centers becomes much more accurate
if several arcs and/or multiple images are present, since in this
case, detailed modelling can be done. This typically proceeds in an
interactive way: First, multiple images have to be identified (based
on their colors and/or detailed morphology, as available with HST
imaging).  Simple (plausible) mass models are then assumed, with
parameters fixed by matching the multiple images, and requiring the
distortion at the arc location(s) to be strong and have the correct
orientation.  This model then predicts the presence of further
multiple images; they can be checked for through morphology and
color. If confirmed, a new, refined model is constructed, which yields
further strong lensing predictions etc.  Such models have predictive
power and can be trusted in quite some detail; the accuracy of mass
estimates in some favourable cases can be as high as a few percent.

In fact, these models can be used to predict the redshift of arcs and
arclets (Kneib et al.\ 1994): since the distortion of a lens depends
on the source redshift, once a detailed mass model is available, one
can estimate the value of the lens strength $\propto D_{\rm ds}/D_{\rm
s}$ and thus infer the redshift. This method has been successfully
applied to HST observations of clusters (Ebbels et al.\ 1998). Of
course, having spectroscopic redshifts of the arcs available increases
the accuracy of the 
calibration of the mass models; they are therefore very useful.

\subsubsection{Results}
The main results of the strong lensing investigations of clusters can
be summarized as follows:
\bi
\item
The mass in cluster centers is much more concentrated than predicted by
(simple) models based on X-ray observations. The latter usually predict a
relatively large core of the mass distribution. These large cores
would render clusters sub-critical to lensing, i.e., they would be
unable to produce giant arcs or multiple images. In fact, when arcs
were first discovered they came as a big surprise because of these
expectations. By now we know that the intracluster medium is much more
complicated than assumed in these `$\beta$-model' fits for the X-ray
emission. 
\item
The mass distribution in the inner part of clusters often shows strong
substructure, or multiple mass peaks. These are also seen in the
galaxy distribution of clusters, but with the arcs can be verified to
also correspond to mass peaks. These are easily understood in the
frame of hierarchical mergers in a CDM model; the merged clusters
retain their multiple peaks for a dynamical time or even longer, and
are therefore not in virial equilibrium.
\item
The orientation of the (dark) matter appears to be fairly strongly
correlated with
the orientation of the light in the cD galaxy; this supports the idea that
the growth of the cD galaxy is related to the cluster as a whole,
through repeated accretion of lower-mass member galaxies. In that
case, the cD galaxy `knows' the orientation of the cluster.
\item
There is in general good agreement between lensing and X-ray mass
estimates for those clusters where a `cooling flow' indicates that they
are in dynamical equilibrium, provided the X-ray analysis takes the
presence of the cooling flow into account.
\ei

\subsection{Mass reconstructions from weak lensing}
Whereas strong lensing probes the mass distribution in the inner part
of clusters, weak lensing can be used to study the mass distribution
at much larger angular separations from the cluster center. In fact,
as we shall see, weak lensing can provide a parameter-free
reconstruction of the projected two-dimensional mass distribution in
clusters. This discovery (Kaiser \& Squires 1993) actually marked the
beginning of quantitative weak lensing research.

\subsubsection{The Kaiser--Squires inversion}
Weak lensing yields an estimate of the local (reduced) shear, as
discussed in Sect.\ 4.2. Here we shall discuss how to derive
the surface mass density from a measurement of the (reduced) shear.
Starting from (\ref{eq:def-pot}) and the definition (\ref{eq:3.12}) of
the shear, one finds that the latter can be written in the form
\begin{eqnarray} 
  \gamma(\vc\theta) &=& \frac{1}{\pi}\int_{\Real^2}\d^2\theta'\,
  {\mathcal D}(\vc\theta-\vc\theta')\,
  \kappa(\vc\theta')    \;,\quad \hbox{with}\nonumber\\
  {\mathcal D}(\vc\theta) &\equiv&
  \frac{\theta_2^2-\theta_1^2-2{\rm i}\theta_1\theta_2}
  {|\vc\theta|^4}
  = \frac{-1}{(\theta_1-{\rm i}\theta_2)^2}  \;.
\elabel{3.15}
\end{eqnarray}
Hence, the complex shear $\gamma$ is a convolution of $\kappa$ with the
kernel $\D$, or, in other words, $\D$ describes the shear generated by
a point mass. In 
Fourier space this convolution becomes a multiplication,
\[
\hat\gamma(\vc
\ell)=\pi^{-1}\hat{\mathcal D}(\vc \ell)\,\hat\kappa(\vc \ell)\quad{\rm for}
\quad \vc \ell\ne \vc 0\;.
\]
This relation can be inverted to yield
\begin{equation}
  \hat\kappa(\vc \ell) = \pi^{-1}\hat\gamma(\vc \ell)\,
  \hat{\mathcal D}^*(\vc \ell)
  \quad\hbox{for}\quad
  \vc \ell\ne\vc 0\;,
\elabel{5.3}
\end{equation}
where
\[
  \hat{\mathcal D}(\vc \ell) = \pi
  \frac{\left(\ell_1^2-\ell_2^2+2{\rm i}\ell_1\ell_2\right)}
  {|\vc \ell|^2}
\]
was used. Fourier back-transformation of (\ref{eq:5.3}) then yields
\be \ebox{
  \kappa(\vc\theta) - \kappa_0 = 
  \frac{1}{\pi}\int_{\Real^2}\d^2\theta'\,
  {\mathcal D}^*(\vc\theta-\vc\theta')\,
  \gamma(\vc\theta')  
  = \frac{1}{\pi}\int_{\Real^2}\d^2\theta'\,
  \Re\left[{\mathcal D}^*(\vc\theta-\vc\theta')\,
  \gamma(\vc\theta')\right] }  \;.
\elabel{5.4}
\ee
Note that the constant $\kappa_0$ occurs since the $\vc\ell=\vc
0$-mode is undetermined. Physically, this is related to the fact that
a uniform surface mass density yields no shear. Furthermore, it is
obvious (physically, though not so easily seen mathematically) that
$\kappa$ must be real; for this reason, the imaginary part of the
integral should be zero, and taking the real- part only makes no
difference. However, in practice it is different, as noisy data, when
inserted into the inversion formula, will produce a non-zero imaginary
part. What (\ref{eq:5.4}) shows is that if $\gamma$ can be measured,
$\kappa$ can be determined.

Before looking at this in more detail, we briefly mention some 
difficulties with the inversion formula as given above:
\bi
\item
Since $\gamma$ can at best be estimated at discrete points (galaxy
images), smoothing is required. One might be tempted to replace the
integral in (\ref{eq:5.4}) by a discrete sum over galaxy positions,
but as shown by Kaiser \& Squires (1993), the resulting mass density
estimator has infinite noise (due to the $\theta^{-2}$-behavior of the
kernel $\D$). 
\item
It is 
not the shear $\gamma$, but the reduced shear $g$ that can be
determined from the galaxy ellipticities; hence, one needs to obtain a
mass density estimator in terms of $g$.
\item
The 
integral in (\ref{eq:5.4}) extends over $\Real^2$, whereas data are
available only on a finite field; therefore, it needs to be seen
whether modifications allow the construction of an estimator for the
surface mass density from 
finite-field shear data.
\item
To get absolute values for the surface mass density,
the additive constant $\kappa_0$ is of course a nuisance. As will be
explained soon, this indeed is the largest problem in mass
reconstructions, and carries the name {\it mass-sheet degeneracy}
(note that we mentioned this effect before, in the context of determining the
Hubble constant from time-delays in lens systems).
\ei

\subsubsection{Improvements and generalizations}
{\bf Smoothing} of data is needed to get a shear field from discrete
data points. When smoothed with Gaussian kernel of angular scale
$\theta_{\rm s}$, the 
covariance of the resulting mass map is finite, and given by (Lombardi
\& Bertin 1998; van Waerbeke 2000)
\[
{\rm Cov}\rund{\kappa(\vc\theta),\kappa(\vc\theta')}
={\sigma_\eps^2\over 4\pi\theta_{\rm s}^2n}
\exp\rund{-{|\vc\theta-\vc\theta'|^2\over 2\theta_{\rm s}^2}}\;.
\]
Thus, the larger the smoothing scale, the less noise does the
corresponding mass map have. Note that, since (i) smoothing can be
represented by a convolution, (ii) the relation between $\kappa$ and
$\gamma$ is a convolution, and (iii) convolution operations are
transitive, it does not matter whether the shear field is smoothed
first and inserted into (\ref{eq:5.4}), or the noisy inversion
obtained by transforming the integral into a sum over galaxy image
positions is smoothed afterwards with the same smoothing kernel.

Noting that it is the reduced shear $g=\gamma/(1-\kappa)$ that can be
estimated from the ellipticity of images, one can write:
\begin{equation}
  \kappa(\vc\theta) - \kappa_0 = 
  \frac{1}{\pi}\,\int_{\Real^2}\d^2\theta'\,
  \left[1-\kappa(\vc\theta')\right]\,\Re\left[
    \D^*(\vc\theta-\vc\theta')\,
    g(\vc\theta')
  \right]\;;
\elabel{5.7}
\end{equation}
this integral equation for $\kappa$ can be solved by iteration, and it
converges quickly. Note that in this case, the
undetermined constant $\kappa_0$ no longer corresponds to adding a
uniform mass sheet. What the arbitrary value of $\kappa_0$ corresponds
to can be seen as follows: 
The transformation
\begin{eqnarray}
  \kappa(\vc\theta)\to\kappa'(\vc\theta) &=&
  \lambda\kappa(\vc\theta)+(1-\lambda)
  \quad\hbox{or}\nonumber\\
  \left[1-\kappa'(\vc\theta)\right] &=&
  \lambda\left[1-\kappa(\vc\theta)\right]
\label{eq:5.8}
\end{eqnarray}
changes the shear $\gamma\to \gamma'=\lambda\gamma$, and thus leaves $g$
invariant; this is the mass-sheet degeneracy!
It 
can be broken if magnification information can be obtained, since
\[
\mu \to \lambda^{-2}\mu \;.
\]
Magnification can in principle be obtained from the number counts of
images (Broadhurst et al.\ 1995), owing to magnification bias,
provided the unlensed number density is sufficiently well
known. Indeed, magnification effects have been detected in a few
clusters as a number depletion of faint galaxy images towards the
center of the clusters (e.g., Fort et al.\ 1997; Taylor et al.\ 1998;
Dye et al.\ 2002).  In principle, the mass sheet degeneracy can also
be broken if redshift information of the source galaxies is available
and if the sources are widely distributed in redshift; however, even
in this case it is only mildly broken, in the sense that one needs a
fairly high number density of background galaxies in order to fix the
parameter $\lambda$ to within $\sim 10\%$.

{\bf Finite-field inversions} start from the relation (Kaiser 1995)
\begin{equation}
  \nabla\kappa = \left(\begin{array}{c}
    \gamma_{1,1}+\gamma_{2,2} \\
    \gamma_{2,1}-\gamma_{1,2} \\
  \end{array}\right) \equiv \vc u_\gamma(\vc\theta)\;,
\elabel{5.10}
\end{equation}
which is a {\it local} relation between shear and surface mass
density; it can easily be derived from the definitions of $\kappa$
(\ref{eq:Poisson}) and $\gamma$ (\ref{eq:3.12}) in terms of
$\psi_{,ij}$. 
A similar relation can be derived in terms of reduced shear,
\be
  \nabla K(\vc\theta) = \frac{-1}{1-g_1^2-g_2^2}\,
  \left(\begin{array}{cc}
    1-g_1 & -g_2 \\
    -g_2 & 1+g_1 \\
  \end{array}\right)\,
  \left(\begin{array}{c}
    g_{1,1}+g_{2,2} \\
    g_{2,1}-g_{1,2} \\
  \end{array}\right) \equiv\vc u_g(\vc\theta)\;,
\label{eq:5.11}
\ee
where
\begin{equation}
  K(\vc\theta) \equiv \ln[1-\kappa(\vc\theta)]
\label{eq:5.12}
\end{equation}
is a non-linear function of $\kappa$.
These equations can be integrated, by formulating them as a von
Neumann boundary-value problem on the data field $\cal U$ (Seitz \&
Schneider 2001),
\begin{equation} \ebox{
  \nabla^2 \kappa =\nabla \cdot \vc u_\gamma
  \quad\hbox{with}\quad
  \vc n\cdot \nabla \kappa = \vc n\cdot \vc u_\gamma
  \quad\hbox{on}\quad
  \partial{\mathcal U}  }  \;;
\label{eq:5.17}
\end{equation}
where $\vc n$ is the outward-directed normal on the boundary
$\partial{\mathcal U}$ of ${\mathcal U}$.
The analogous equation holds for $K$ in terms of $g$ and $\vc
u_g$. The numerical solution of these equations is 
fast, using overrelaxation (see Press et al.\ 1992).
In fact, the foregoing
formulation of the problem is equivalent (Lombardi \& Bertin 1998)
to the minimization of the action
\be
  A=\int_{\mathcal U}\d^2\theta\; |\nabla\kappa(\vc\theta)-\vc
   u_\gamma(\vc\theta)|^2 \;,
\label{eq:5.17a}
\end{equation}
from which the von Neumann problem can be derived as the Euler
equation of the variational principle $\delta A=0$. 
These parameter-free mass reconstructions have been applied to quite a
number of clusters; it provides a tool to make their dark matter
distribution `visible'.

\subsubsection{Results}
\begin{figure}
\centerline{\psfig{figure=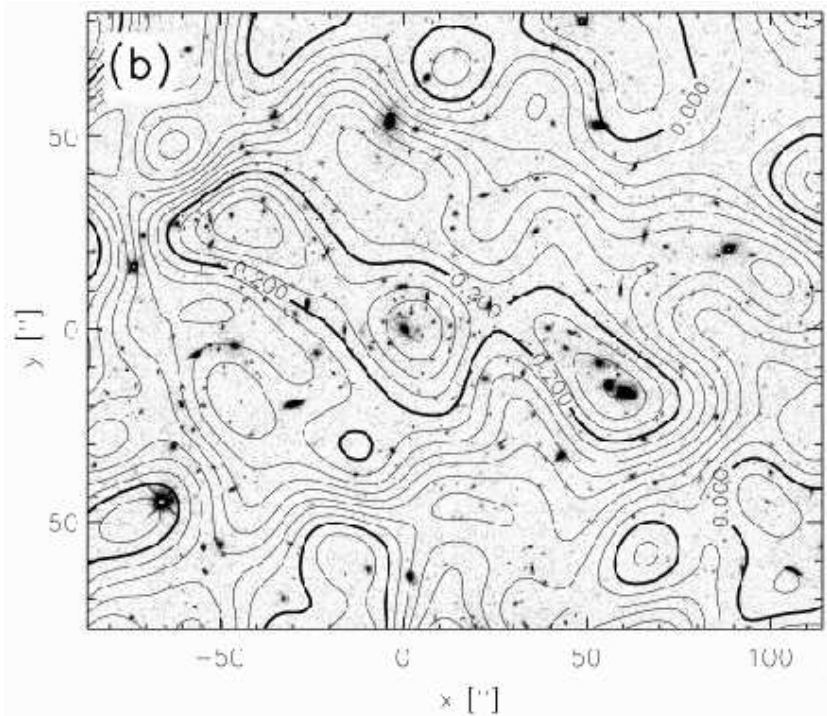,width=11cm}}
\caption{Mass reconstruction (contours) of the inner part of the high redshift
($z_{\rm d}=0.83$) cluster MS1054$-$03, based on a mosaic of six
pointings obtained with the WFPC2@HST (from Hoekstra et al.\
2000). The splitting of the cluster core into three subcomponents, 
previously seen from ground-based images by Clowe et al.\ (2000),
shows that this cluster is not yet relaxed}
\label{fig:1054}
\end{figure}
The mass reconstruction techniques discussed above have been applied
to quite a number of clusters up to now, yielding parameter-free mass
maps. It is obvious that the quality of a mass map depends on the
number density of galaxies that can be used for a shear estimate,
which in turn depends on the depth and the seeing of the observational
data. Furthermore, the mass profiles of clusters are much more
reliably determined if the data field covers a large region, as
boundary effects get minimized. 

The first application of the Kaiser \& Squires reconstruction
technique was done to the X-ray detected cluster MS1224+20 (Fahlman et
al.\ 1994); it resulted in an estimate of the mass-to-light ratio in
this $z=0.33$ cluster of $\sim 800\,h$, considerably larger than
`normal' values of $M/L\sim 250\, h$. This conclusion was later
reinforced by a fully independent weak-lensing analysis of this
cluster by Fischer (1999). This mass estimate is in fact in conflict
with the measured velocity dispersion of the cluster galaxies, which
is much smaller than obtained by an SIS fit to the shear data. The
line-of-sight to this cluster is fairly complicated, with additional
peaks in the redshift distribution of galaxies in the field (Carlberg
et al.\ 1994), all of which are included in the weak lensing
measurement.  Furthermore, this cluster may not be in a relaxed state,
which probably renders the X-ray mass analysis inaccurate.

In fact, non-relaxed clusters are probably more common than naively
expected. One example is shown in Fig.\ \ref{fig:1054}, a mass
reconstruction of a high-redshift cluster based on HST data. The
presence of three mass clumps, which coincide with three
concentrations of cluster galaxies, indicates that this cluster is
still in the process of merging. When the merging occurs along the
line-of-sight, then it is less obvious in the mass maps. One example
seems to be the cluster Cl0024+16, for which one obtains a large mass
from the distance of its arcs from the cluster center (Colley et al.\
1996), but which is fairly underluminous in X-rays for this mass. A
detailed investigation of the structure of this cluster in radial
velocity space by Czoske et al.\ (2002) has shown strong evidence for
a collision of two clusters along the line-of-sight. Another example
is provided by the cluster A1689, where the extended arc structures
suggest an Einstein radius of about $40''$ for this cluster, making
this the strongest lensing cluster in the sky, but the weak lensing
results do not support the enormous mass obtained from the arcs (see
Clowe \& Schneider 2001, King et al.\ 2002, and references therein).

However, in many clusters the weak lensing mass estimates are in good
agreement with those from dynamical estimates and X-ray determinations
(e.g., Squires et al.\ 1996), provided the inner region of the
clusters are omitted -- but for them, the weak lensing method does not
have sufficient angular resolution anyway. For example, Hoekstra
and collaborators have observed three X-ray selected clusters with HST
mosaics, and their results, summarized in Hoekstra et al.\ (2002a),
shows that the SIS fit values for the velocity dispersion agree with
those from spectrosopic investigations.

The mass maps can also be used to study how well the cluster galaxy
distribution traces the underlying dark matter. An HST data based mass
reconstruction of Cl0939+47 (Seitz et al.\ 1996) shows detailed
structure that is very well matched with the distribution of bright
cluster galaxies. A more quantitative investigation was performed by 
Wilson et al.\ (2001) showing that early-type galaxies trace the dark
matter distribution quite well.

One of the predictions of CDM models for structure formation is that
clusters of galaxies are located at the intersection points of
filaments. In particular, this implies that a physical pair of
clusters should be connected by a bridge or filament of (dark) matter,
and weak lensing mass reconstructions can in principle be used to
search for them. In the investigation of the $z=0.42$ supercluster
MS0302, Kaiser et al.\ (1998) found an indication of a possible
filament connecting two of the three clusters, with the caveat (as
pointed out by the authors) that the filament lies just along the
boundary of two CCD chips. Gray et al.\ (2002) saw a filament
connecting the two clusters A901A/901B in their mass recosntruction of
the A901/902 supercluster field. One of the problems related to the
unambiguous detection of filaments is the difficulty to define what a
`filament' is, i.e. to device a statistics to quantify the presence of
a mass bridge. Because of that, it is difficult to distinguish between
noise in the mass maps, the `elliptical' extension of two clusters
pointing towards each other, and a true filament. 

A perhaps surprising result is the difficulty of distinguishing
between the NFW mass profile from, say power-law models, such as the
isothermal profile. In fact, even from weak lensing observations
covering large fields, out to the virial radius of clusters (Clowe \&
Schneider 2001, 2002; King et al.\ 2002), the distinction between NFW
and isothermal is present only at the $\lesssim 2\sigma$ level. The
reason for this is the mass-sheet degeneracy. Within a family of
models (such as the NFW), the model parameters can be determined with
fairly high accuracy. One way to improve on the distinction between
various mass profiles is to incorporate strong lensing constraints
into the mass reconstruction (e.g., using inverse methods; see
below). In particular, the multiple images seen in the cores of
clusters can be used to determine the central mass profile (see Sand
et al.\ 2002; Gavazzi et al.\ 2003). On the other hand, one can
statistically superpose weak lensing measurements of clusters to
obtain their average mass profile, as done by Dahle et al.\ (2003) for
six clusters; even in that case, a (generalized) NFW profile is hardly
distinguishable from an isothermal model.

\subsubsection{Inverse methods}
In addition to these `direct' methods for determing $\kappa$, inverse
methods have been developed, such as a maximum-likelihood fit
(Bartelmann et al.\ 1996) to the data. In these techniques, one
parameterizes the lens by the deflection potential $\psi$ on a grid
and then maximizes
\begin{equation}
 {\cal L} = \prod_{i=1}^{N_{\rm g}}{1\over
 \pi\,\sigma_i^2\rund{\vc\theta_i,\{\psi_n\}}} 
 \exp\rund{ - 
 \frac{|\epsilon_i-g\rund{\vc\theta_i,\{\psi_n\}}|^2}
{\sigma_i^2\rund{\vc\theta_i,\{\psi_n\}}} }
 \elabel{5.35}
\end{equation}
with respect to these gridded $\psi$-values. In order to avoid
overfitting, one needs a regularization; entropy regularization (Seitz
et al.\ 1998) seems best suited. It should be pointed out that the
deflection potential $\psi$, and not the surface mass density
$\kappa$, should be used as a variable, for two reasons: first, shear
and $\kappa$ depend locally on $\psi$, and are thus readily calculated
by finite differencing, whereas the relation between $\gamma$ and
$\kappa$ is non-local and requires summation over all
gridpoints. Second, and more importantly, the surface mass density on a
finite field {\it does not} determine $\gamma$ on this field, since
mass outside the field contributes to $\gamma$ as well.

There are a number of reasons why inverse methods are in principle
preferable to the direct method discussed above. First, in the direct
methods, the smoothing scale is set arbitrarily, and in general kept
constant. It would be useful to have an objective way how to choose 
this scale, and perhaps, the smoothing scale be a function
of position: e.g., in regions with larger number densities of sources,
the smoothing scale could be reduced. Second, the direct methods do
not allow additional input coming from observations; for example, if
both shear and magnification information are available, the latter
could not be incorporated into the mass reconstruction. The same is
true for clusters where strong lensing constraints are known.

\subsection{Aperture mass}
In the weak lensing regime, $\kappa\ll 1$, the mass-sheet degeneracy
corresponds to adding a uniform surface mass density $\kappa_0$. We
shall now consider a quantity, linearly related to $\kappa$, that is
unaffected by the mass-sheet degeneracy. Let
$U\rund{|\vc\theta|}$ be
a compensated weight (or filter) function, with 
$\int \d\theta\;\theta\,U(\theta)=0$,
then the {\it aperture mass}
\begin{equation}
  M_{\rm ap}(\vc\theta_0) = \int\d^2\theta\,\kappa(\vc\theta)\,
  U(|\vc\theta-\vc\theta_0|)
\elabel{5.22}
\end{equation}
is independent of $\kappa_0$, as can be easily seen. The important
point to notice is that $M_{\rm ap}$ can be
written directly in terms of the shear (Schneider 1996)
\begin{equation} \ebox{
  M_{\rm ap}(\vc\theta_0) = \int\d^2\theta\,{Q(|\vc\theta|)}\,
  \gamma_{\rm t}(\vc\theta;\vc\theta_0) } \;,
\label{eq:5.28}
\end{equation}
where we have defined the {\em tangential component\/}
$\gamma_{\rm t}$ of the shear relative to the point
$\vc\theta_0$, and 
\be
Q(\theta)={2\over \theta^2}\int_0^\theta\d\theta'\;\theta'\,
U(\theta') - U(\theta)\;.
\ee
These relations can be derived from (\ref{eq:5.10}), by rewriting the
partial derivatives in polar coordinates and subsequent integration by parts.

We shall now consider a few properties of the aperture mass.
\bi
\item
If $U$ has finite support, then $Q$ has finite support. This implies
that the aperture mass can be calculated on a finite data field.
\item
If $U(\theta)={\rm const.}$ for $0\le \theta\le\theta_{\rm in}$, then 
$Q(\theta)=0$ for the same interval. Therefore,
the strong lensing regime (where the shear $\gamma$ deviates
significantly from the reduced shear $g$) can be
avoided by properly choosing $U$ (and $Q$).
\item 
If $U(\theta)=(\pi \theta_{\rm in}^2)^{-1}$ for
$0\le\theta\le \theta_{\rm in}$, 
$U(\theta)=-[\pi (\theta_{\rm out}^2-
\theta_{\rm in}^2)]^{-1}$ for $\theta_{\rm in} < \theta\le \theta_{\rm
out}$, and $U=0$ for $\theta>\theta_{\rm out}$,
then 
$Q(\theta)=\theta_{\rm out}^2\,\theta^{-2}\left[\pi
(\theta_{\rm out}^2-\vartheta_{\rm in}^2)\right]^{-1}$ for
$\theta_{\rm in}\le\theta\le\theta_{\rm out}$, 
and $Q(\theta)=0$
otherwise. For this special choice of $U$, 
\be
M_{\rm ap}=\bar\kappa(\theta_{\rm in})-\bar\kappa(\theta_{\rm
in},\theta_{\rm out})\;,
\ee
the mean mass density inside $\theta_{\rm in}$ minus the mean density
in the annulus $\theta_{\rm in}\le\theta\le\theta_{\rm out}$.
Since the latter is non-negative, this yields lower limit to
$\bar\kappa(\theta_{\rm in})$, and thus to $M(\theta_{\rm in})$.
\ei

\setcounter{equation}{0}
\section{Cosmic shear -- lensing by the LSS}
Up to now we have considered the lensing effect of localized mass
concentrations, like galaxies and clusters. In addition to that, light
bundles propagating through the Universe are continuously deflected
and distorted by the gravitational field of the inhomogeneous mass
distribution, the large-scale structure (LSS) of the cosmic matter
field (the reader is referred to John Peacock's lecture for the
definition of cosmological parameters and the theory of structure
growth in the Universe).
This distortion of light bundles causes shape distortions of
images of distant galaxies, and therefore, the statistics of the
distortions reflect the statistical properties of the LSS.

{\em Cosmic shear} deals with the investigation of this connection,
from the measurement of the correlated image distortion to the
inference of cosmological information from this distortion
statistics. As we shall see, cosmic shear has become a very important
tool in observational cosmology. From a technical point-of-view, it is
quite challenging, first because the distortions are indeed very weak
and therefore difficult to measure, and second, in contrast to
`ordinary' lensing, here the light deflection does not occur in a
`lens plane' but by a 3-D matter distribution; one therefore
needs a different description of the lensing optics. We start by
looking at the description of light propagating through the Universe.

\subsection{Light propagation in an inhomogeneous Universe}
The laws of light propagation follow from Einstein's General
Relativity; according to it, light propagates along the
null-geodesics of the space-time metric. As shown in SEF, one can
derive from General Relativity that 
the governing equation for the propagation of thin light bundles
through an arbitrary space-time is the equation of geodesic deviation,
\begin{equation}
  \frac{\d^2\vc\xi}{\d\lambda^2} = {\mathcal T}\,\vc\xi\;,
\label{eq:6.1}
\end{equation}
where $\vc\xi$ is the separation vector of two neighboring light rays,
$\lambda$ the affine parameter along the central ray of the bundle,
and ${\mathcal T}$ is the {\em optical tidal matrix\/} which describes
the influence of space-time curvature on the propagation of light.
${\mathcal T}$ can be expressed directly in terms of the Riemann
curvature tensor.  

For the case of a weakly inhomogeneous Universe, the tidal matrix can
be explicitly calculated in terms of the Newtonian potential.  For
that, we write the slightly perturbed metric of the Universe in the
form
\begin{equation}
  \d s^2 = a^2(\tau)\left[
    \left(1+\frac{2\Phi}{c^2}\right)\,c^2\d\tau^2-
    \left(1-\frac{2\Phi}{c^2}\right)
    \left(\d w^2+f_K^2(w)\d\omega^2\right)
  \right]\;,
\label{eq:3.38}
\end{equation}
where $w$ is the comoving radial distance, $a=(1+z)^{-1}$ the scale
factor, normalized to unity today, $\tau$ is the conformal time,
related to the cosmic time $t$ through $\d t=a\;\d \tau$, $f_K(w)$ is
the comoving angular diameter distance, which equals $w$ in a
spatially flat model, and $\Phi$ denotes the Newtonian peculiar
gravitational potential. In this metric, the equation of geodesic
deviation yields, for the comoving separation vector $\vc
x(\vc\theta,w)$ between a ray separated by an angle $\vc\theta$ at the
observer from a fiducial ray, the evolution equation
\begin{equation} 
\frac{\d^2\vc x}{\d w^2} +
K\,\vc x = -\frac{2}{c^2} \eck{ \nabla_\perp\Phi\rund{\vc
x(\vc\theta,w),w} -\nabla_\perp\Phi^{(0)}\rund{w}} \;,
\label{eq:6.12}
\end{equation}
where $K=(H_0/c)^2\,(\Omega_{\rm m}+\Omega_\Lambda-1)$ is the spatial
curvature, $\nabla_\perp=(\partial/\partial x_1,\partial/\partial
x_2)$ is the transverse {\it comoving} gradient operator, and
$\Phi^{(0)}(w)$ is the potential along the fiducial ray.  The formal
solution of the transport equation is obtained by the method of
Green's function, to yield
\begin{equation}
  \vc x(\vc\theta,w) = f_K(w)\vc\theta - 
  \frac{2}{c^2}\int_0^w\! \d w'\,f_K(w-w') 
  \eck{ \nabla_\perp\Phi\rund{\vc
x(\vc\theta,w'),w'} -\nabla_\perp\Phi^{(0)}\rund{w'}} \;.
\label{eq:6.15}
\end{equation}
A source at comoving distance $w$ with comoving separation $\vc x$
from the fiducial light ray would be
seen, in the absence of lensing, at the angular separation $\vc \beta=\vc
x/f_K(w)$ from the fiducial ray (this statement is nothing but the
definition of the comoving angular diameter distance).
Hence, in analogy with standard lens theory, we define the Jacobian matrix
\be
\A(\vc\theta,w)={\partial\vc \beta\over\partial\vc\theta}
={1\over f_K(w)}{\partial\vc x\over\partial\vc\theta}\;,
\ee
and obtain
\be
\A_{ij}(\vc\theta,w)=\delta_{ij}-\frac{2}{c^2}\int_0^w \!\! \d w'\,
{f_K(w-w') f_K(w')\over f_K(w)}\,\Phi_{,ik}\rund{\vc
x(\vc\theta,w'),w'}\,\A_{kj}(\vc\theta,w') \;,
\ee
which describes the locally linearized mapping introduced by LSS
lensing.  This equation still is exact in the limit of validity of the
weak-field metric. Next, we expand $\A$ in powers of $\Phi$, and
truncate the series after the linear term:
\be \ebox{
\A_{ij}(\vc\theta,w)=\delta_{ij}-\frac{2}{c^2}\int_0^w \!\! \d w'\,
{f_K(w-w') f_K(w')\over f_K(w)}\,\Phi_{,ij}\rund{f_K(w')\vc\theta,w'} } \;.
\ee
Hence, to linear order, the distortion can be obtained by integrating
along the unperturbed ray; this is also called the Born
approximation. Corrections to the Born approximation are necessarily
of order $\Phi^2$.
If we now define the deflection potential
\be
\psi(\vc\theta,w):=\frac{2}{c^2}\int_0^w \!\! \d w'\,
{f_K(w-w') f_K(w')\over f_K(w)}\,\Phi\rund{f_K(w')\vc\theta,w'}
\ee
then $\A_{ij}=\delta_{ij}-\psi_{,ij}$, just as in ordinary lens theory.
{\em In this approximation, lensing by the 3-D matter
distribution can be treated as an equivalent lens plane with
deflection potential $\psi$, mass density $\kappa=\nabla^2\psi/2$, and
shear $\gamma=(\psi_{,11}-\psi_{,22})/2+{\rm i}\psi_{,12}$. }

\subsection{Cosmic shear: the principle}
\subsubsection{The effective surface mass density}
Next, we relate $\kappa$ to the fractional density contrast $\delta$ of
matter fluctuations in the Universe; this is done in a number of
steps:
\ben
\item
Take the 2-D Laplacian of $\psi$, and add the term $\Phi_{,33}$ in the
integrand; this latter term vanishes in the line-of-sight integration,
as can be seen by integration by parts.
\item
We make use of the 3-D Poisson equation in comoving coordinates
\be
\nabla^2\Phi={3 H_0^2\Omega_{\rm m}\over 2 a}\delta
\ee
to obtain
\be \ebox{
\kappa(\vc\theta,w)=\frac{3H_0^2\Omega_{\rm m}}{2c^2}\,
  \int_0^w\,\d w'\,\frac{f_K(w')f_K(w-w')}{f_K(w)}\,
  \frac{\delta\rund{f_K(w')\vc\theta,w'}}{a(w')} } \;.
\label{eq:6.21}
\end{equation}
Note that $\kappa$ is proportional to $\Omega_{\rm m}$, since lensing is
sensitive to $\Delta\rho\propto \Omega_{\rm m}\,\delta$, not just to
the density contrast $\delta=\Delta\rho/\bar\rho$ itself.
\item
For a redshift distribution of sources with $p_z(z)\,\d z=p_w(w)\,\d w$,
the effective surface mass density becomes
\be \ebox{
\kappa(\vc\theta)=\int\d w\;p_w(w)\,\kappa(\vc\theta,w)=
\frac{3H_0^2\Omega_{\rm m}}{2c^2}\,
  \int_0^{w_{\rm h}}\d w\;g(w)\,f_K(w)\,
\frac{\delta\rund{f_K(w)\vc\theta,w}}{a(w)} }
\elabel{5.11a}
\ee
with
\be \ebox{
g(w)=\int_w^{w_{\rm h}}\d w'\;p_w(w'){f_K(w'-w)\over f_K(w')} } \;,
\ee
which is essentially the source-redshift weighted lens efficiency factor
$D_{\rm ds}/D_{\rm s}$ for a
density fluctuation at distance $w$, and
$w_{\rm h}$ is the comoving horizon distance.
\een

\subsubsection{Limber's equation}
The density field $\delta$ is assumed to be a realization of a random
field. It is the properties of the random field that cosmologists can
predict. In particular, the second-order statistical
properties of the density field are described in terms of the power
spectrum. We shall therefore look at the relation between the quantities
relevant for lensing and the power spectrum. The basis of this
relation is Limber's equation.
If $\delta$ is a homogeneous and isotropic 3-D random field, then the
projections 
\be
g_i(\vc\theta)=\int\d w\;q_i(w)\,\delta\rund{f_K(w)\vc\theta,w}
\ee
also are (2-D) homogeneous and isotropic random fields, where the
$q_i$ are weight functions. In particular, the correlation function
\be
C_{12}=\ave{g_1(\vc\vp_1)\,g_2(\vc\vp_2)}\equiv
C_{12}(|\vc\vp_1-\vc\vp_2|) 
\ee
depends only on the modulus of the separation vector. The original
form of the Limber equation relates $C_{12}$ to the correlation
function of $\delta$ which is a line-of-sight
projection. Alternatively, one can consider the Fourier-space version
of this relation: The power spectrum $P_{12}(\ell)$ -- the Fourier
transform of $C_{12}(\theta)$ -- depends linearly on the power
spectrum $P_\delta(k)$ of the density fluctuations (Kaiser 1992), 
\be\ebox{
P_{12}(\ell)=\int\d w \;{q_1(w)\, q_2(w)\over f_K^2(w)}\,
P_\delta\rund{{\ell\over f_K(w)},w} } \;,
\elabel{limber}
\ee
if the largest-scale structures in $\delta$ are much smaller than the
effective range $\Delta w$ of the projection.
Hence, we obtain the (very reasonable) result that the power at
angular scale $1/\ell$ is obtained from the 3-D power at
length scale $f_K(w)\,(1/\ell)$, integrated over $w$. Comparing
(\ref{eq:5.11}) with (\ref{eq:limber}), one sees that 
$\kappa(\vc\theta)$ is such a projection of $\delta$ with
the weights
$q_1(w)=q_2(w)= (3/2)(H_0/c)^2\Omega_{\rm m}g(w)f_K(w)/a(w)$, so that
\begin{equation} \ebox{
  P_\kappa(\ell) = \frac{9H_0^4\Omega_{\rm m}^2}{4c^4}\,
  \int_0^{w_{\rm h}}\d w\,\frac{g^2(w)}{a^2(w)}\,
  P_\delta\left(\frac{\ell}{f_K(w)},w\right)  } \;.
\label{eq:6.25}
\end{equation}
The power spectrum $P_\kappa$, if obtained through observations, can
therefore be used to constrain the 3-D power spectrum $P_\delta$.

\bc
\ec

\subsection{Second-order cosmic shear measures}
As we shall see, all second-order statistics of the cosmic shear yield
(filtered) information about $P_\kappa$.
The most-often used second-order statistics are:
\bi
\item
The two-point correlation function(s) of the shear, $\xi_{\pm}(\theta)$,
\item
the shear dispersion in a (circular) aperture,
$\ave{|\bar\gamma|^2}(\theta)$, and 
\item
the aperture mass dispersion, $\ave{M_{\rm ap}^2}(\theta)$.
\ei
These will be discussed next, and their relation to $P_\kappa(\ell)$
shown. As a preparation, consider the Fourier transform of $\kappa$, 
\be
\hat\kappa(\vc\ell)=\int\d^2\theta\,{\rm e}^{{\rm
i}\vc\ell\cdot\vc\theta}\,\kappa(\vc\theta)\;;
\ee
then,
\be
\ave{\hat\kappa(\vc\ell)\hat\kappa^{*}(\vc\ell')}
=(2\pi)^2\,\delta_{\rm D}(\vc\ell-\vc\ell')\,P_\kappa(\ell)\;,
\ee
which provides another definition of the power spectrum $P_\kappa$. 
The Fourier transform of the shear is
\be
\hat\gamma(\vc\ell)=\rund{\ell_1^2-\ell_2^2+2{\rm i}\ell_1\ell_2
\over \abs{\vc\ell}^2}\hat\kappa(\vc\ell)
\elabel{N10}
\ee
which implies that
\be
\ave{\hat\gamma(\vc\ell)\hat\gamma^*(\vc\ell')}
=(2\pi)^2\,\delta_{\rm D}(\vc\ell-\vc\ell')\,P_\kappa(\ell).
\ee
Hence, the power spectrum of the shear is the same as that of the
convergence.

\subsubsection{Shear correlation functions}
Consider a pair of points (i.e., galaxy images); their separation direction
$\vp$ (i.e. the polar angle of the separation vector $\vc\theta$) 
is used to define the tangential and cross-component of the
shear at these positions {\em for this pair},
\be
\gamma_{\rm t}=-\Re\rund{\gamma\,{\rm e}^{-2{\rm i}\vp}} \;,\quad
\gamma_{\times}=-\Im\rund{\gamma\,{\rm e}^{-2{\rm i}\vp}} \;.
\ee
Then, the shear correlation functions are defined as  
\bea
\xi_\pm(\theta)&=&\ave{\gamma_{\rm t}\gamma_{\rm t}} \pm\ave{\gamma_\times
\gamma_\times}(\theta)\;,\nonumber\\
\xi_\times(\theta)&=&\ave{\gamma_{\rm t}\gamma_{\times}}(\theta)
\;.\nonumber 
\eea
Due to parity symmetry, $\xi_\times(\theta)$ is expected to vanish, since
under such a transformation, $\gamma_{\rm t}\to \gamma_{\rm t}$, but
$\gamma_\times\to -\gamma_\times$.
Next we relate the shear correlation functions to the power spectrum
$P_\kappa$:
Using the definition of $\xi_\pm$, replacing $\gamma$ in terms of
$\hat\gamma$, and making use of relation between $\hat\gamma$ and
$\hat\kappa$, one finds:
\be \ebox{
\xi_+(\theta)=\int_0^\infty
{\d\ell\,\ell\over 2\pi}\,{\rm J}_0(\ell\theta)\,
P_\kappa(\ell)\;;  \;\;
\xi_-(\theta)=\int_0^\infty
{\d\ell\,\ell\over 2\pi}\,{\rm J}_4(\ell\theta)\,
P_\kappa(\ell)  }  \;.
\ee
$\xi_\pm$ can be measured as follows: on a data field, select all
pairs of faint galaxies with separation within $\Delta\theta$ of
$\theta$ and then
take the average $\ave{\eps_{{\rm t}i}\,\eps_{{\rm t}j}}$ over all
these pairs; since $\eps=\eps^{(\rm s)}+\gamma(\vc\theta)$, the expectation
value of $\ave{\eps_{{\rm t}i}\,\eps_{{\rm t}j}}$ is 
$\ave{\gamma_{\rm t}\gamma_{\rm t} }(\theta)$, provided source
ellipticities are uncorrelated. 
Similarly, the correlation for the cross-components is obtained.

\subsubsection{The shear dispersion}
Consider a circular aperture of radius $\theta$; the mean shear in this
aperture is $\bar\gamma$. Averaging over many such apertures, one
defines the shear dispersion
$\ave{|\bar\gamma|^2}(\theta)$.
It is related to the power spectrum through
\be \ebox{
\ave{\abs{\bar\gamma}^2}(\theta)={1\over 2\pi}\int\d\ell\,\ell\,
P_\kappa(\ell)\,W_{\rm TH}(\ell\theta) } \;,\;\;
{\rm where} \;\;
W_{\rm TH}(\eta)={4 {\rm J}_1^2(\eta)\over \eta^2}
\ee
is the top-hat filter function. The shear dispersion
can be measured by averaging the square of the mean galaxy
ellipticities over many independent apertures.

\subsubsection{The aperture mass}
Consider a circular aperture of radius $\theta$; for a point inside the
aperture, define the tangential and cross-components of the shear relative
to center of aperture (as before);
then define
\be
M_{\rm ap}(\theta)=\int\d^2\vt\;Q(|\vc\vt|)\,\gamma_{\rm t}(\vc\vt)\;,
\ee
where $Q$ is a weight function with support $\vt\in[0,\theta]$.
In the following we shall use
\[
Q(\vt)={6\over \pi\theta^2}\,{\vt^2\over\theta^2}
\rund{1-{\vt^2\over\theta^2}}\,{\rm H}(\theta-\vt)\;,
\]
in which case the dispersion of $M_{\rm ap}(\theta)$ is related to
the power spectrum by
\be \ebox{
\ave{M_{\rm ap}^2}(\theta)={1\over
2\pi}\int_0^\infty\d\ell\;\ell\,P_\kappa(\ell)\, W_{\rm
ap}(\theta\ell) }  \;,\;\;
{\rm with} \;\;
W_{\rm ap}(\eta):={576{\rm J}_4^2(\eta)\over \eta^4}\;.
\elabel{N26a}
\ee
\subsubsection{Interrelations}
These various 2-point statistics all depend linearly on the power
spectrum $P_\kappa$; therefore, one should not be too surprised that
they are all related to each other. The surprise perhaps is that these
interrelations are quite simple (Crittenden et al.\ 2002).
First, the relations between $\xi_\pm$ and $P_\kappa$ can be inverted,
making use of the orthonormality relation of Bessel functions:
\be
P_\kappa(\ell)=2\pi\int_0^\infty\d\theta\,\theta\,\xi_+(\theta)\,{\rm
J}_0(\ell\theta)=2\pi\int_0^\infty\d\theta\,\theta\,\xi_-(\theta)\,{\rm
J}_4(\ell\theta) \;.
\elabel{Pfromxi}
\ee
Next, we 
take one of these and plug it into the relation between the other
correlation function and $P_\kappa$, to find:
\be \ebox{
\xi_+(\theta)=\xi_-(\theta)
+\int_\theta^\infty{\d\vt\over\vt}\xi_-(\vt)\rund{4-12{\theta^2\over
\vt^2}} } \;;
\ee
\be\ebox{
\xi_-(\theta)=\xi_+(\theta)
+\int_0^\theta{\d\vt\,\vt\over\theta^2}\xi_+(\vt)
\rund{4-12{\vt^2\over \theta^2}} } \,.
\ee
Using (\ref{eq:Pfromxi}) in the expression for the shear dispersion,
one finds 
\be \ebox{
\ave{\abs{\bar\gamma}^2}(\theta)=
\int_0^{2\theta}{\d\vt\,\vt\over  \theta^2}
\;\xi_+(\vt)\,S_+\rund{\vt\over\theta}
=
\int_0^\infty{\d\vt\,\vt\over  \theta^2}
\;\xi_-(\vt)\,S_-\rund{\vt\over\theta}  } \;,
\ee
where the $S_\pm$ are simple functions, given explicitly in Schneider
et al.\ (2002a). Finally, the same procedure for the aperture mass
dispersion lets us write
\be \ebox{
\ave{M_{\rm ap}^2}(\theta)=\int_0^{2\theta}{\d\vt\,\vt\over  \theta^2}
\;\xi_+(\vt)\,T_+\rund{\vt\over\theta} 
=
\int_0^{2\theta}{\d\vt\,\vt\over  \theta^2}
\;\xi_-(\vt)\,T_-\rund{\vt\over\theta} } \;,
\ee
again with known functions $T_\pm$ (Schneider et al.\ 2002a).
Hence, all these 2-point statistics can be evaluated from the
correlation functions $\xi_\pm(\theta)$, which is of particular
interest, since they  can be measured best: 
Real data fields contain holes and gaps (like CCD defects, brights stars,
nearby galaxies, etc.) which makes the placing of apertures difficult;
however, the 
evaluation of the correlation functions is not affected by gaps, as
one uses all pairs of galaxy images with a given angular separation. 

\subsection{Cosmic shear and cosmology}
\subsubsection{Why cosmology from cosmic shear?}
Before continuing, it is worth to pause for a second and ask the question
as to why one tries to investigate cosmological questions by using cosmic
shear -- since the CMB can measure cosmological
parameters with high accuracy. Partial answers to this question are:
\bi
\item
Cosmic shear measures the mass distribution at much lower redshifts
($z\lesssim 1$) and at smaller physical scales [$R\sim 0.3\,h^{-1}\,
(\theta/1')\,{\rm Mpc}$] than the CMB; indeed, it is the only way to
map out the dark matter distribution directly without any assumptions
about the relation between dark and baryonic matter. The smaller
scales probed are very important for constraining the shape of the
power spectrum, i.e., the primordial tilt and the shape parameter
$\Gamma_{\rm spect}$. 
\item
Cosmic shear measures the non-linearly evolved mass
distribution and its associated power spectrum $P_\delta(k)$; hence, in
combination with the CMB it allows us to study the evolution of the
power spectrum and in particular, provides a very powerful test of the
gravitational instability paradigm for structure growth.
\item
It provides a fully independent way to probe the cosmological model.
Given the revolutionary claims coming from the CMB, SN\ Ia, and the
LSS of the galaxy distribution, namely that more than 95\% of the
stuff in the Universe is in a form about whose physical nature we have
not the slightest idea, an additional independent verification of these
claims is certainly welcome. 
\item
As we shall see shortly, cosmic shear studies provide a new and
highly valuable search method for cluster-scale matter
concentrations. 
\ei
\subsubsection{Expectations}
The cosmic shear signal depends on the cosmological model,
parametrized by $\Omega_{\rm m}$, $\Omega_\Lambda$, and the shape
parameter $\Gamma_{\rm spect}$ of the power spectrum, the
normalization of the power spectrum, usually expressed in terms of
$\sigma_8$, and the redshift distribution of the sources.  By
measuring $\xi_\pm$ over a significant range of angular scales one can
derive constraints on these parameters.

The accuracy with which $\xi_\pm$ can be measured depends on number
density of galaxies (that is, depth and quality of the images), the
total solid angle covered by the survey, and its geometric arrangement
(compact survey vs. widely separated pointings); it is
determined by a combination of intrinsic ellipticity dispersion and
the cosmic (or sampling) variance. For angular scales below about 1
degree, the non-linear evolution of the power spectrum becomes
important for the cosmic shear signal; because of this, the expected
signal is considerably larger than estimated from linear perturbation
theory of structure evolution. Furthermore, the signal depends quite
strongly on the mean redshift of the source galaxies, which suggests
that deep surveys, aiming for higher-redshift galaxies, are best
suited for cosmic shear studies.

\subsection{Observation of cosmic shear}
\subsubsection{First detections}
Whereas the theory of cosmic shear was worked out in the early 1990's
(Blandford et al.\ 1991; Miralda-Escud\'e 1991; Kaiser 1992), it took
until the year 2000 before this effect was first discovered. 
The reason for this time lag must be seen as a combination of
instrumental developments, i.e.\ the wide-field CCD mosaic cameras,
and the image analysis software, like IMCAT, with which shapes of
galaxies can be corrected for PSF effects. Finally, in March 2000, four
groups independently published their first discoveries of cosmic shear
(Bacon et al.\ 2000; Kaiser et al.\ 2000; van Waerbeke et al.\ 2000,
Wittman et al.\ 2000). In these surveys, of the order of $10^5$ galaxy
images have been analyzed, covering about 1\ deg$^2$. The 
fact that the results from these four independent teams agreed within
the respective error bars immediately gave credence to this new window
of observational cosmology. Furthermore, 4
different telescopes, 5 different cameras, independent data reduction
tools and at least two different image analysis methods have been
used in these studies. Maoli et al.\ (2001) reported a significant cosmic shear
measurement from 50 widely separated FORS1@VLT images, which also
agreed with the earlier measurements.

\subsubsection{Deriving constraints}
From the measured correlation functions $\xi_\pm(\theta)$, obtaining
constraints on cosmological parameters can proceed through minimizing
\be
\chi^2(p)=\sum_{ij}\rund{\xi_i(p)-\xi_i^{\rm obs}}\,{\rm
Cov}^{-1}_{ij} \,\rund{\xi_j(p)-\xi_j^{\rm obs}}\;,
\ee
with $\xi_i=\xi(\theta_i)$ being the binned correlation function(s)
(i.e., either $\xi_\pm$, or using both), $p$ is a set of cosmological
parameters, and ${\rm Cov}^{-1}_{ij}$ the inverse covariance
matrix. The latter can be determined either from the $\xi_\pm$ itself,
from simulations, or estimated from the data (see Schneider et al.\
2002b). Nevertheless, the calculation of the covariance is fairly
cumbersome, and most authors have used approximate methods to derive
it, such as the field-to-field variations of the measured correlation.
As it turns out, $\xi_+ (\theta)$ is strongly correlated
across angular bins, much less so for $\xi_-(\theta)$; this is due to
the fact that the filter function that describes $\xi$ in terms of the
power spectrum $P_\kappa$ is much broader for $\xi_+$ (namely ${\rm
J}_0$) than ${\rm J}_4$ which applies for $\xi_-$. Of course, a
corresponding figure-of-merit function can be defined for the other
second-order shear statistics, with their respective covariance
matrices, but as argued before, the correlation functions should be
regarded as the basic observable statistics.

\subsubsection{Recent results}
Since the first detections of cosmic shear, described above, there
have been a large number of measurements over the past three
years. Instead of mentioning them all here, we refer the reader to
the recent reviews by van Waerbeke \& Mellier (2003) and Refregier
(2003). State-of-the-art are deep surveys, similar to those with which
the first cosmic shear results have been derived, but with significantly larger
solid angle (van Waerbeke et al.\ 2001, 2002), or shallower surveys of
much larger area (e.g., Hoekstra et al.\ 2002c; Jarvis et al.\ 2003).
The results of these surveys, which contain of the order of $\sim
10^6$ galaxies, i.e., an order-of-magnitude more than the discovery
surveys mentioned above, 
can be summarized roughly as follows:

\begin{figure}
\centerline{\psfig{figure=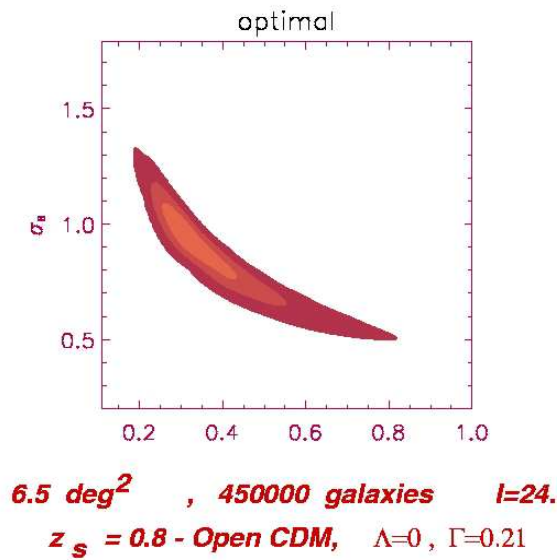,width=9cm}}
\caption{Constraints in the $\Omega_{\rm m}$ -- $\sigma_8$ parameter
plane, from the VIRMOS-DESCART survey (van Waerbeke et al.\
2001). For this figure, in which the 1, 2, and 3-$\sigma$ confidence
regions are indicated,
a zero cosmological constant has been assumed,
the redshift distribution of the source galaxies was assumed to be
known, as well as the shape parameter $\Gamma_{\rm spect}=0.21$}
\flabel{vW01}
\end{figure}

Cosmic shear by itself presently does not provide strong constraints
on multi-dimensional cosmological parameter space. Hence, if one does
not fix most of the cosmological parameters from external sources, the
allowed region in multi-dimensional parameter space is still quite
large. On the other hand, if one considers a restricted set of
parameters, cosmic shear results are very powerful.  An example of
that is given in Fig.\ \ref{fig:vW01}, where all cosmological
parameters have been kept fixed, except $\Omega_{\rm m}$ and the
normalization $\sigma_8$. In this case, one finds indeed a
well-defined maximum of the corresponding likelihood. Interestingly,
the direction of the `likelihood valley' nearly coincides with the
constraint obtained from the cluster abundance, i.e., it provides a
constraint on $\sigma_8\Omega_{\rm m}^{0.6}$. If the other
cosmological parameters are not assumed to be known precisely, but are
marginalized over a plausible uncertainty range, the likelihood
contours widen substantially. 

When combined with results from other methods, cosmic shear yields
very useful information. For example, as pointed out by Hu \& Tegmark
(1999), when combined with data from CMB anisotropy, cosmic shear can
break degeneracies of model parameters which are present when using
the CMB data alone. In the $\Omega_{\rm m}$ -- $\sigma_8$ parameter
plane, cosmic shear constraints are nearly perpendicular to those from
the CMB (van Waerbeke et al.\ 2002).

Therefore, at present the best use of cosmic shear results is in
constraining the normalization $\sigma_8$ of the density
perturbations, for a set of other cosmological parameters fixed by
other methods, such as the CMB, galaxy redshift surveys, etc. The
various cosmic shear surveys have given a range of $\sigma_8$
determinations which is about as narrow as current estimates from the
abundance of massive clusters (see van Waerbeke \& Mellier 2003 for a
summary of these results). Given the youth of this field, this indeed
is a remarkable achievement already. Furthermore, since the
determination of $\sigma_8$ from cluster abundance and cosmic shear
agree, one learns something important: the cluster abundance depends
on the assumed Gaussianity of the primordial density field, whereas
the constraint from cosmic shear does not. Hence, the agreement
between the two methods supports the idea of an initial Gaussian
field. Without doubt, the next generation of
cosmic shear surveys will provide highly accurate determinations of
this normalization, as well as other (combinations of) cosmological
parameters.

\subsection{E-modes, B-modes}
In the derivation of the lensing properties of the LSS, we ended up
with an equivalent surface mass density.  In particular, this implied
that $\A$ is a symmetric matrix, that the shear can be obtained in
terms of $\kappa$ or $\psi$. Now, the shear is a 2-component quantity,
whereas both $\kappa$ and $\psi$ are scalar fields. This implies that
the two shear components are not independent of each other!

Recall that (\ref{eq:5.10}) yields a relation between the gradient of
$\kappa$ and the first derivatives of the shear components; in
particular, (\ref{eq:5.10}) implies that $\nabla\times \vc
u_\gamma\equiv 0$, yielding a local constraint relation between the
shear components. The validity of this constraint equation guarantees
that the imaginary part of (\ref{eq:5.4}) vanishes.  This constraint
is also present at the level of 2-point statistics, since one expects
from (\ref{eq:Pfromxi}) that
\be
\int_0^\infty\d\theta\,\theta\,
\xi_+(\theta){\rm J}_0(\theta\ell)
=
\int_0^\infty\d\theta\,\theta\,
\xi_-(\theta){\rm J}_4(\theta\ell)\;.
\ee
Hence, the two correlation functions $\xi_\pm$ are not independent.
The observed shear field is not guaranteed to satisfy these relations,
due to noise, remaining systematics, or other effects. Therefore,
searching for deviations from this relation allows a check for these
effects.  However, there might also be a `shear' component present
that is not due to lensing (by a single equivalent thin matter sheet
$\kappa$). Shear components which satisfy the foregoing relations are called
E-modes; those which don't are B-modes -- these names are exported
from the polarization of the CMB, which has the same mathematical
properties as the shear field.

The best way to separate these modes locally is provided by the
aperture measures: 
$\ave{M^2_{\rm ap}(\theta)}$ is sensitive {\it only} to E-modes.
If one defines in analogy
\be
M_{\perp}(\theta)=\int\d^2\vt\;Q(|\vc\vt|)\,\gamma_{\times}(\vc\vt)\;,
\elabel{Mapperp}
\ee
then $\ave{M^2_\perp(\theta)}$ is sensitive {\it only} to B-modes.

\begin{figure}
\centerline{\psfig{figure=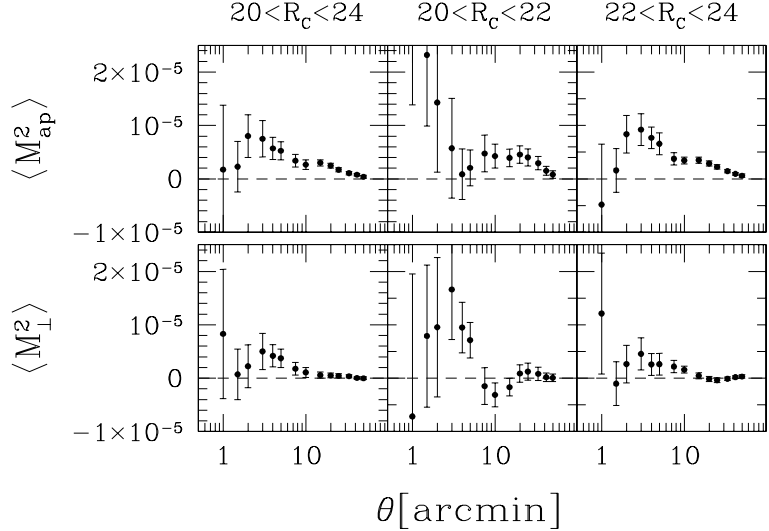,width=11cm}}
\caption{Dispersion of the aperture mass $\ave{M_{\rm ap}^2}$ (upper
row) and its analogue $\ave{M_\perp^2}$ for the cross-component, as
obtained from the Red Cluster Sequence survey (Hoekstra et al.\ 2002).
The left panels show the results for a broad range of galaxy
brightnesses, whereas the middle and right rows display the results for
the bright and the fainter parts, respectively, of the
sample. Clearly, the presence of the B-mode is seen; its strength
decreases for the fainter part of the sample; this behavior is
expected if the B-mode is due to intrinsic alignments of galaxies. Its
relative importance decreases with increasing width of the redshift
distribution of galaxies} 
\flabel{hoek}
\end{figure}

Significant B-modes have been discovered in cosmic shear surveys
(e.g., van Waerbeke et al.\ 2002; Hoekstra et al.\ 2002 -- see Fig.\
\ref{fig:hoek}); the question now is what are they due to? As
mentioned before, the noise, which contributes to both E- and B-modes
in similar strengths, could be underestimated, there could be
remaining systematic effects, or indeed show the real presence of a
B-mode on the sky. There are two possibilities known to generate a
B-mode through 
lensing: The first-order in $\Phi$ (or `Born') approximation may not
be strictly valid, but as shown by ray-tracing simulations through
cosmic matter fields (e.g., Jain et al.\ 2000) the resulting B-modes
are expected to be very small. Clustering of sources also yields a
finite B-mode (Schneider et al.\ 2002a), but again, this effect is
much smaller than the observed amplitude of the B-modes.

Currently the best guess for the generation of a finite B-mode are
intrinsic correlations of galaxy ellipticities.  Such intrinsic
alignments of galaxy ellipticities can be caused by the tidal
gravitational field of the large-scale structure at
galaxy formation.  Predictions of the alignment of the projected
ellipticity of the galaxy mass can be made analytically (e.g. tidal torque
theory) or from numerical simulations; however, the predictions from
various groups differ by large factors (e.g., Croft \& Metzler 2000;
Crittenden et al.\ 2001; Heavens et al.\ 2000; Jing 2002) which means
that the process is not well understood at present. In addition, there
remains the question of whether the orientation of the galaxy light
(which is the issue of relevance here) is the same as that of the
mass.

If intrinsic alignments play a role, then
\be
\xi_+=\ave{\eps_i\,\eps^*_j}=\ave{\eps_i^{(\rm s)}\,\eps^{(\rm s)*}_j}
+\xi_+^{\rm lens}\;,
\ee
and measured correlations $\xi_\pm$ contain both components.  Of
course, there is no reason why intrinsic correlations should have only
an B-mode. If a B-mode contribution is generated through this process,
then the measured E-mode is also contaminated by intrinsic alignments.
In fact, the various models do not agree on the relative strength of
E- and B-modes in the intrinsic alignments of galaxies, but it seems
that the E-modes have generally higher amplitude than the B-modes. 
Given that intrinsic alignments yield ellipticity correlations only for
spatially close sources (i.e., close in 3-D, not merely in
projection), it is clear that the deeper a cosmic shear survey is, and
thus the broader the redshift distribution, the
smaller is the relative amplitude of an intrinsic signal. Most of the
theoretical predictions on the strength of intrinsic alignments say
that the deep cosmic shear surveys (say, with mean source redshifts of
$\ave{z_{\rm s}}\sim 1$) are affected at a $\sim 10\%$
level, but that shallow cosmic shear surveys are more strongly
affected; for them, the intrinsic alignment can
be of the same order as, or larger than the lensing signal. 

However, the intrinsic signal can be separated from the lensing signal
if redshift information of the sources is available, owing to the fact
that $\ave{\eps_i^{(\rm s)}\,\eps^{(\rm s)*}_j}$ will be non-zero only if
the two galaxies are at the same redshift. Hence, 
if $z$-information is available (e.g., photometric redshifts), then galaxy
pairs which are likely to have similar redshifts are to be avoided in
estimating the cosmic shear signal (King \& Schneider 2002; Heymans \&
Heavens 2003). This will change the expectation value of the shear
correlation function, but in a controlable way, as the redshifts are
assumed to be known. Indeed, using (photometric) redshifts, one can
simultaneously determine the intrinsic and the lensing signal,
essentially providing a cosmic shear tomography (King \& Schneider
2003). 

\subsection{Higher-order statistics}
On the level of second-order statistics, `only' the power spectrum is
probed. If the density field was Gaussian, then the power spectrum
would fully characterize it; however, in the course of non-linear
structure evolution, non-Gaussian features of the density field are
generated, which show up correspondingly in the cosmic shear field and
which can be probed by higher-order shear statistics.  The usefulness
of these higher-order measures for cosmic shear has been pointed out
in Bernardeau et al.\ (1997) and van Waerbeke et al.\ (1999); in
particular, the near-degeneracy between $\sigma_8$ and $\Omega_{\rm
m}$ can be broken.  However, these are serious problems with
higher-order shear statistics, that shall be illustrated in terms of
the third-order statistics. The three-point correlation function has
three independent variables (e.g. the sides of a triangle) and 8
components; as was shown in Schneider \& Lombardi (2003), none of
these eight components vanishes owing to parity invariance. This then
implies that the covariance matrix has 6 arguments and 64 components!
Of course, this is too difficult to handle efficiently, and therefore
one must 
ask which combinations of the components of the 3-pt correlation
function are most useful for studying the dark matter distribution.
Unfortunately, this is essentially unknown yet. An additional problem
is that the predictions from theory are less well established than for
the second-order statistics.

Nevertheless, progress has been made. From ray-tracing simulations
through a cosmic matter distribution, the 3-pt correlation function of
the shear can be determined (Takada \& Jain 2003; see also Zaldarriaga
\& Scoccimarro 2003); in addition, Schneider \& Lombardi (2003) have defined
the `natural components' of the 3-pt correlator which are most easily
related to the bispectrum of the underlying matter distribution.

Alternatively, aperture measures can be defined to measure the
third-order statistics. Schneider et al.\ (1998) calculated
$\ave{M_{\rm ap}^3}(\theta)$ in the frame of the quasi-linear
structure evolution model and showed it to be a strong function of
$\Omega_{\rm m}$. Indeed, $\ave{M_{\rm ap}^3}$ is sensitive only to
the E-modes of the shear field. One might be tempted to use
$\ave{M_\perp^3}(\theta)$ as a measure for third-order B-mode
statistics, but indeed, this quantity vanishes owing to parity
invariance. However, $\ave{M_\perp^2\,M_{\rm ap}}$ is a measure for
the B-modes at the third-order statistical level. Bernardeau et al.\
(2002) measured for the first time a significant 3-rd order shear from
the VIRMOS-DESCART survey, employing a suitably filtered integral over
the measured 3-pt correlation function. With the upcoming large cosmic
shear surveys, the 3-pt function will be measured with high accuracy.

\subsection{Weak lensing search for cluster-mass dark halos}
As we have seen, the mass distribution of clusters of galaxies can be
mapped by weak lensing techniques. In fact, the coherent alignment of
background galaxy images clearly shows the presence of a massive
matter concentration present at or near the location of the optically
or X-ray selected cluster towards which the weak lensing observations
were targeted. As pointed out in Schneider (1996), one can use weak
lensing to search for clusters: seeing a strong alignment of galaxy
images 
centered onto a point on a wide-field image, one would conclude the
presence of the mass concentration there. A very useful way to
quantify this is the aperture mass statistics, already introduced. By
selecting an appropriate filter function, one can systematically seach
for statistically significant peaks of $M_{\rm ap}$ on wide-field
images. In fact, the data needed for this investigation is the same as
that used in cosmic shear surveys.

Since clusters of galaxies are very important cosmological probes,
e.g., to determine the normalization of the power spectrum of the
matter inhomogeneities in the Universe, a selection of clusters based
on their mass properties only would be extremely useful. Usually,
clusters are selected by their optical or X-ray properties; to
transform luminosity or X-ray temperature into a mass estimate, and
thus to transform a flux-limited cluster sample into a mass-limited
sample, which can then be compared to cosmological predictions, one
needs to employ a number of approximations and scaling relations. In
contrast to this, the shear selection can directly be compared to
cosmological predictions, e.g., by calculating the abundance of peaks
of $M_{\rm ap}$ directly from N-body simulations of structure
formation (e.g., Reblinsky et al.\ 1999), without reliance on
the luminous properties of baryonic matter, nor even for
identifying cluster-mass halos in the simulated density fields.

The abundance of peaks above a given threshold $M_{\rm ap}$, at a
given angular scale, can also be used as a cosmic shear measure. In
fact, in future large cosmic shear surveys this will become most
likely one of the most useful statistics for studying non-Gaussian
aspects of the shear field. Within the frame of Press--Schechter
theory, Kruse \& Schneider (1999) calculated the $M_{\rm ap}$ peak
statistics, which was then compared with direct numerical simulation
by Reblinsky et al.\ (1999). White et al.\ (2002) pointed out that the
$M_{\rm ap}$ statistics can be substantially affected by the
large-scale structure along the line-of-sight to these mass
concentrations; this implies that the relation between $M_{\rm ap}$
and the mass of the clusters is not simple -- but again, this method
does not require a mass function to be determined, as the $M_{\rm
ap}$-statistics can be obtained directly from LSS simulations. 

Several clusters, or cluster candidates, have been found that
way. Erben et al.\ (2000) detected a highly significant shear signal
corresponding to a putative mass peak, about $7'$ away from the
cluster Abell\ 1942, seen on two images taken with different filters
and different cameras. No obvious concentration of galaxies is seen in
this direction, neither in the optical nor near-IR images (Gray et
al.\ 2001), making it a candidate for a `dark clump'; however, before
drawing this conclusion, further investigations are needed, such as
imaging with the HST to confirm the shear measurements. Umetsu \&
Futamase (2000) found a significant mass concentration on an HST
image, again without an obvious optical counterpart. In contrast to this,
Mellier et al.\ (2000) found a mass peak in one of their 50 VLT fields
taken for a cosmic shear survey with FORS, which is clearly associated
with a concentration of galaxies. Wittman et al.\ (2001, 2002)
detected two clusters on their wide-field images, and confirmed them
spectroscopically. In fact, making use of photometric redshift
estimates of the background galaxies, and employing the redshift
dependence of the lens strength, they were able to estimate rather
precisely the cluster redshifts, which were later confirmed with
spectroscopy. Two of the three putative mass concentrations found by
Dahle et al.\ (2002) are also very likely to be associated with
luminous clusters. Hence, the shear selection of clusters has already
been proven as a very useful concept.

\subsection{Lensing in three dimensions}
Using (photometric) redshift estimates in cosmic shear research is not
only useful to remove the potential contribution from intrinsic
alignments of galaxy ellipticities. If one defines galaxy populations
with different redshift distributions, one can probe different
projections of the cosmic density field; see
(\ref{eq:5.11}). This then increases the information one can extract
from a cosmic shear survey, and thus the ability to discriminate
between different cosmological models (Hu 1999).

More recently, it was pointed out (Taylor 2001; Hu \& Keeton 2002)
that the use of redshift information can in principle be used to
reconstruct the three-dimensional density field $\delta$ from the shear
measurements. This is based on the possibility to invert
(\ref{eq:5.11}), i.e., to express $\delta(w)$ in terms of
$\kappa(w)$. The study of the 3-dimensional mass distribution is
particularly interesting for constraining the properties of the dark
energy in the Universe (e.g., Heavens 2003). 

\section{Conclusions}
Due to its insensitivity to the nature of matter causing the
gravitational potential, gravitational lensing has turned out to be an
ideal tool to probe the structure of the (dark) matter distribution in
the Universe, from small to large scales. Progress in this field has
been very rapid in the past years, and due to the fast pace at which
new instruments become available, it is guaranteed to continue its role
as an important tool for observational cosmology. For example, the
first images from the new camera ACS are breathtaking and will
certainly lead to much improved mass models of clusters of
galaxies. The new square degree optical cameras will provide cosmic
shear surveys covering an appreciable fraction of the sky -- as a
consequence, statistical uncertainties and cosmic variance will then
no longer be the main contribution to the error budget, but systematic
effects in estimating shear from CCD images may well take over. The
Next Generation Space Telescope will provide a superb tool for
studying galaxy-scale lens systems, as well as clusters. 

\section*{Acknowledgement}
I would like to thank John Beckmann for the invitation to this School;
being there was a very nice experience. A big thanks to him and
his colleagues for the great organization of the School and the
associated cultural events.  The active participation of the students
and their curiosity was very much appreciated. I would like to thank
my colleagues and friends with whom I had the pleasure
to discuss the topics of this article over the years, in particular Matthias
Bartelmann, Doug Clowe, Thomas Erben, Lindsay King (special thanks to
her for carefully reading and commenting this manuscript), Marco
Lombardi, Yannick Mellier, Ludovic van Waerbeke, my past and current
students, as well as Chris Kochanek and Joachim Wambsganss with whom I
had the fun to teach (and learn!) for a week at the 2003 Saas-Fee
course.  This work was supported by the German Ministry for Science
and Education (BMBF) through the DLR under the project 50 OR 0106, by
the German Ministry for Science and Education (BMBF) through DESY
under the project 05AE2PDA/8, and by the Deutsche
Forschungsgemeinschaft under the project SCHN 342/3--1.


\begin{thebibliography}{} 
\bibitem[2000]{BRE00}
  Bacon, D.J., Refregier, A.R. \& Ellis, R.S. 2000, MNRAS, 318, 625
\bibitem[1]{1}
Bacon, D.J., Refregier, A., Clowe, D. \& Ellis, R.S.\ 2001, MNRAS 325,
1065
\bibitem[1]{1}
Bartelmann, M.\ 1995, A\&A 299, 11
\bibitem[1]{1}
Bartelmann, M., Narayan, R., Seitz, S. \& Schneider, P.\ 1996,
ApJ 464, L115
\bibitem[1]{1}
Bartelmann, M. \& Schneider, P.\ 2001, Physics Reports 340, 291
\bibitem[1]{1}
Bernardeau, F., Mellier, Y. \& van Waerbeke, L.\ 2002, A\&A 389, L28
\bibitem[1]{1}
Bernardeau, F., Van Waerbeke, L. \& Mellier, Y. 1997, A\&A, 322, 1
\bibitem[1]{1}
Bernstein, G.M. \& Jarvis, M.\ 2002, AJ 123, 583
\bibitem[1]{1}
Binney, J. \& Tremaine, S.\ 1987, {\it Galactic dynamics}, Princeton
University Press, Princeton.
\bibitem[1]{1}
Blandford, R.D., Saust, A.B., Brainerd, T.G. \& Villumsen,
  J.V. 1991, MNRAS, 251, 600
\bibitem[1]{1}
Bonnet, H. \& Mellier, Y.\ 1995, A\&A 303, 331
\bibitem[1]{1}
Bradac, M., Schneider, P., Steinmetz, M., Lombardi, M., King, L.J. \&
Porcas, R.\ 2002, A\&A 388, 373
\bibitem[1]{1}
Bradac, M., Schneider, P., Lombardi, M.\ et al.\ 2003, astro-ph/0306238
\bibitem[1]{1}
Brainerd, T.G., Blandford, R.D. \& Smail, I.\ 1996, ApJ 466, 623
\bibitem[1]{1}
Broadhurst, T.J., Taylor, A.N. \& Peacock, J.A.\ 1995, ApJ 438, 49
\bibitem[1]{1}
Burke, W.L.\ 1981, ApJ 244, L1
\bibitem[1]{1}
Carlberg, R.G., Yee, H.K.C. \& Ellingson, E.\ 1994, ApJ 437, 63
\bibitem[1]{1}
Catelan, P., Kamionkowski, M. \& Blandford, R.D.\ 2001, MNRAS, 320, L7
\bibitem[1]{1}
Clowe, D., Luppino, G.A., Kaiser, N. \& Gioia, I.M.\ 2000 ApJ 539, 540
\bibitem[1]{1}
Clowe, D. \& Schneider, P.\ 2001, A\&A 379, 384
\bibitem[1]{1}
Clowe, D. \& Schneider, P.\ 2002, A\&A 395, 385
\bibitem[1]{1}
Colley, W.N., Tyson, J.A. \& Turner, E.L.\ 1996, ApJ 461, L83
\bibitem[1]{1}
Crittenden, R.G., Natarajan, P., Pen, U.-L. \& Theuns, T. 2001, ApJ,
  559, 552
\bibitem[1]{1}
Crittenden, R.G., Natarajan, P., Pen, U.-L. \& Theuns, T. 2002,
  ApJ, 568, 20
\bibitem[1]{1}
Croft, R.A.C. \& Metzler, C.A.\ 2001, ApJ, 545, 561
\bibitem[1]{1}
Czoske, O., Moore, B., Kneib, J.-P. \& Soucail, G.\ 2002, A\&A 386, 31
\bibitem[1]{1}
Dahle, H., Pedersen, K., Lilje, P.B., Maddox, S.J. \& Kaiser, N.\
2002, astro-ph/0208050
\bibitem[1]{1}
Dahle, H., Hannestad, S. \& Sommer-Larsen, J.\ 2003, ApJ 388, L73
\bibitem[1]{1}
Dalal, N. \& Kochanek, C.S.\ 2002, ApJ 572, 25
\bibitem[1]{1}
Dye, S., Taylor, A.N., Greve, T.R.\ et al.\ 2002, A\&A 386, 12
\bibitem[1]{1}
Ebbels, T., Ellis, R., Kneib, J.-P.\ et al.\ 1998, MNRAS 295, 75
\bibitem[1]{1}
Erben, T., van Waerbeke, L., Mellier, Y., Schneider, P., Cuillandre,
J.C., Castander, F.J. \& Dantel-Fort, M.\ 2000, A\&A 355, 23
\bibitem[1]{1}
Erben, T., van Waerbeke, L., Bertin, E., Mellier, Y. \& Schneider, P.\
2001, A\&A 366, 717
\bibitem[1]{1}
Fahlman, G., Kaiser, N., Squires, G. \& Woods, D.\ 1994, ApJ 437, 56
\bibitem[1]{1}
Falco, E.E., Gorenstein, M.V. \& Shapiro, I.I.\ 1985, ApJ 289, L1
\bibitem[1]{1}
Falco, E.E., Impey, C.D., Kochanek, C.S.\ et al.\ 1999, ApJ 523, 617
\bibitem[1]{1}
Fischer, P.\ 1999, AJ 117, 2024
\bibitem[1]{}
Fischer, P., McKay, T.A., Sheldon, E.\ et al.\ 2000, AJ 120, 1198
\bibitem[1]{1}
Fort, B. \& Mellier, Y.\ 1994, A\&A Review 5, 239
\bibitem[1]{1}
Fort, B., Mellier, Y. \& Dantel-Fort, M.\ 1997, A\&A 321, 353
\bibitem[1]{1}
Freedman, W.L., Madore, B.F., Gibson, B.K.\ et al.\ 2001, ApJ 553, 47
\bibitem[1]{1}
Gavazzi, R., Fort, B., Mellier, Y., Pell\'o, R. \& Dantel-Fort, M.\
2003, A\&A 403, 11
\bibitem[1]{1}
Gray, M.E., Ellis, R.S., Lewis, J.R., McMahon, R.G. \& Firth, Andrew
E.\ 2001, MNRAS 325, 111
\bibitem[1]{1}
Gray, M.E., Taylor, A.N., Meisenheimer, K., Dye, S., Wolf, C. \&
Thommes, E.\ 2002, ApJ 568, 141
\bibitem[1]{}
Guzik, J. \& Seljak, U.\ 2002, MNRAS 335, 311
\bibitem[1]{1}
Heavens, A.F.\ 2003, astro-ph/0304151
\bibitem[1]{1}
Heavens, A.F., Refregier, A. \& Heymans, C.E.C. 2000, MNRAS, 319, 649
\bibitem[1]{1}
Heymans, C. \& Heavens, A.\ 2003, MNRAS 339, 711
\bibitem[1]{1}
Hoekstra, H., Franx, M. \& Kuijken, K.\ 2000, ApJ 532, 88
\bibitem[1]{1}
Hoekstra, H., Franx, M., Kuijken, K. \& van Dokkum, P.G.\ 2002a, MNRAS
333, 911
\bibitem[1]{1}
Hoekstra, H., van Waerbeke, L., Gladders, M.D., Mellier, Y. \& Yee,
H.K.C.\ 2002b, ApJ 577, 604
\bibitem[1]{1}
Hoekstra, H., Yee, H.K.C. \& Gladders, M.D.\ 2002c, ApJ 577, 595
\bibitem[1]{1}
Hu, W.\ 1999, ApJ 522, L21
\bibitem[1]{1}
Hu, W. \& Keeton, C.R.\ 2002, astro-ph/0205412
\bibitem[1]{1}
Hu, W. \& Tegmark, M.\ 1999, ApJ 514, L65
\bibitem[1]{1}
Jain, B., Seljak, U. \& White, S.D.M. 2000, ApJ, 530, 547
\bibitem[1]{1}
Jarvis, M., Bernstein, G.M., Fischer, P.\ et al.\ 2003, AJ 125, 1014
\bibitem[1]{1}
Jing, Y.P.\ 2002, MNRAS 335, L89
\bibitem[1]{1}
Kaiser, N.\ 1992, ApJ, 388, 272
\bibitem[1]{1}
Kaiser, N.\ 1995, ApJ 439, L1
\bibitem[1]{1}
Kaiser, N.\ 2000, ApJ 537, 555
\bibitem[1]{1}
Kaiser, N., Squires, G., 1993, ApJ, 404, 441
\bibitem[1]{1}
Kaiser, N., Squires, G. \& Broadhurst, T.\ 1995, ApJ 449, 460
\bibitem[1]{1}
Kaiser, N., Wilson, G, Luppino, G.\ et al.\ 1998, astro-ph/9809268
\bibitem[2000]{KWL00}
  Kaiser, N., Wilson, G. \& Luppino, G. 2000, astro-ph/0003338
\bibitem[1]{1}
Keeton, C.R., Kochanek, C.S. \& Falco, E.E.\ 1998, ApJ 509, 561
\bibitem[1]{1}
King, L.J., Jackson, N., Blandford, R.D.\ et al.\ 1998, MNRAS 289, 450
\bibitem[1]{1}
King, L., Clowe, D.I. \& Schneider, P.\ 2002, A\&A 383, 118
\bibitem[1]{1}
King, L. \& Schneider, P.\ 2002, A\&A 396, 411
\bibitem[1]{1}
King, L. \& Schneider, P.\ 2003, A\&A 398, 23
\bibitem[1]{1}
Kneib, J.-P., Mathez, G., Fort, B., Mellier, Y., Soucail, G. \&
Longaretti, P.-Y.\ 1994, A\&A 286, 701
\bibitem[1]{1}
Kochanek, C.S.\ 1995, ApJ 445, 559
\bibitem[1]{1}
Kochanek, C.S.\ 2002, ApJ 578, 25
\bibitem[1]{1}
Kochanek, C.S.\ 2003, ApJ 583, 49
\bibitem[1]{1}
Kochanek, C.S. \& Dalal, N.\ 2003, astro-ph/0302036
\bibitem[1]{1}
Kuijken, K.\ 1999, A\&A 352, 355
\bibitem[1]{1}
Koopmans, L.V.E. \& Treu, T.\ 2003, ApJ 583, 606
\bibitem[1]{1}
Kruse, G. \& Schneider, P.\ 1999, MNRAS 302, 821
\bibitem[1]{1}
Langston, G.I., Conner, S.R., Lehar, J., Burke, B.F. \& Weiler, K.W.\
1990, Nature 344, 43
\bibitem[1]{1}
Lombardi, M. \& Bertin, G.\ 1998, A\&A 335, 1
\bibitem[1]{}
Luppino, G.A. \& Kaiser, N.\ 1997, ApJ 475, 20
\bibitem[1]{}
Mao, S. \& Schneider, P.\ 1998, MNRAS 295, 587
\bibitem[1]{}
McKay, T.A., Sheldon, E., Racusin, J.\ et al.\ 2001, astro-ph/0108013
\bibitem[1]{1}
Mellier, Y. et al.\ 2000, ESO Messenger 102, 30
\bibitem[1]{1}
Mellier, Y.\ 1999, ARA\&A 37, 127
\bibitem[1]{1}
Miralda-Escud\'e, J. 1991, ApJ 380, 1
\bibitem[2001]{maoli01}
  Maoli, R., van Waerbeke, L., Mellier, Y., et al.\ 2001, A\&A, 368, 766
\bibitem[1]{1}
Narayan, R. \& Bartelmann, M.\ 1999, in: {\it Formation of Structure
in the Universe}, A.\ Dekel \& J.P.\ Ostriker (eds.), Cambridge
University Press, p.360 
\bibitem[1]{1}
Navarro, J.F., Frenk, C.S. \& White, S.D.M.\ 1997, ApJ 490, 493
\bibitem[1]{1}
Paczy\'nski, B.\ 1996, ARA\&A 34, 419
\bibitem[1]{1}
Petters, A.O., Levine, H. \& Wambsganss, J.\ 2001, Singularity Theory
and Gravitational Lensing (Boston: Birkh\"auser)
\bibitem[1]{1}
Press, W.H., Flannery, B.P., Teukolsky, S.A. \& Vetterling, W.T.\
1986, Numerical Recipes, (Cambridge: University Press)
\bibitem[1]{1}
Reblinsky, K., Kruse, G., Jain, B. \& Schneider, P.\ 1999, A\&A 351, 815
\bibitem[1]{1}
Refregier, A.\ 2003, ARA\&A, submitted
\bibitem[1]{1}
Refregier, A. \& Bacon, D.\ 2003, MNRAS 338, 48
\bibitem[1]{1}
Roulet, E. \& Mollerach, S.\ 1997, Physics Report 279, 67
\bibitem[1]{1}
Rusin, D., Kochanek, C.S., Falco, E.E.\ et al.\ 2003, ApJ 587, 143
\bibitem[1]{1}
Sand, D.J., Treu, T. \& Ellis, R.S.\ 2002, ApJ 574, L129
\bibitem[1]{1}
Schechter, P.L. \& Wambsganss, J.\ 2002, ApJ 580, 685
\bibitem[1]{1}
Schmidt, R.W., Kundic, T., Pen, U.-L.\ et al.\ 2002, A\&A 392, 773
\bibitem[1]{1}
Schneider, P.\ 1984, A\&A 140, 119
\bibitem[1]{1}
Schneider P., 1996, MNRAS, 283, 83
\bibitem[1]{1}
Schneider P., 1998, ApJ 498, 43
\bibitem[1]{1}
Schneider, P., Ehlers, J. \& Falco, E.E.\ 1992, Gravitational Lenses
(New York: Springer)
\bibitem[1]{1}
Schneider, P. \& Lombardi, M.\ 2003, A\&A 397, 809
\bibitem[1]{1}
Schneider, P. \& Seitz, C.\ 1995, A\&A 294, 411
\bibitem[1]{1}
Schneider P., van Waerbeke L., Jain B., Kruse G., 1998, MNRAS, 296,
  873 
\bibitem[1]{1}
Schneider, P., van Waerbeke, L. \& Mellier, Y.\ 2002a, A\&A 389, 729
\bibitem[1]{1}
Schneider, P., van Waerbeke, L., Kilbinger, M. \& Mellier, Y.\ 2002b,
A\&A 396, 1
\bibitem[1]{1}
Schramm, T. \& Kayser, R.\ 1995, A\&A 299, 1
\bibitem[1]{1}
Seitz, C., Kneib, J.-P., Schneider, P. \& Seitz, S.\  1996, A\&A 314, 707
\bibitem[1]{1}
Seitz, C. \& Schneider, P.\ 1997, A\&A 318, 687
\bibitem[1]{1}
Seitz, S. \& Schneider, P.\ 1992, A\&A 265, 1
\bibitem[1]{1}
Seitz, S. \& Schneider, P.\ 2001, A\&A, 374, 740
\bibitem[1]{1}
Seitz, S., Schneider, P. \& Bartelmann, M.\ 1998, A\&A 337, 325
\bibitem[1]{1}
Squires, G., Kaiser, N., Babul, A.\ et al.\ 1996, ApJ 461, 572
\bibitem[1]{1}
Takada, M. \& Jain, B.\ 2003, MNRAS 340, 580
\bibitem[1]{1}
Taylor, A.N.\ 2001, astro-ph/0111605
\bibitem[1]{1}
Taylor, A.N., Dye, S., Broadhurst, T.J., Ben{\'\i}tez, N. \& van
Kampen, E.\ 1998, ApJ 501, 539
\bibitem[1]{1}
Treu, T. \& Koopmans, L.V.E.\ 2002, MNRAS 337, L6
\bibitem[1]{1}
Umetsu, K. \& Futamase, T.\ 2000, ApJ 539, L5
\bibitem[1]{1}
van Waerbeke, L. 1998, A\&A 334, 1
\bibitem[1]{1}
van Waerbeke, L. 2000, MNRAS, 313, 524
\bibitem[1]{1}
van Waerbeke, L., Bernardeau, F. \& Mellier, Y.\ 1999, A\&A, 243, 15
\bibitem[2000]{vWetal00}
  Van Waerbeke, L., Mellier, Y., Erben, T.\ et al. 2000, A\&A, 358, 30
\bibitem[1]{1}
van Waerbeke, L. \& Mellier, Y.\ 2003, astro-ph/0305089
\bibitem[1]{1}
van Waerbeke, L., Mellier, Y., Pell\'o, R., Pen, U.-L., McCracken,
H.J. \& Jain, B.\ 2002, A\&A 393, 369
\bibitem[1]{1}
Wallington, S., Kochanek, C.S. \& Narayan, R.\ 1996, ApJ 465, 64
\bibitem[1]{1}
Wambsganss, J.\ 1998, Living Reviews in Relativity, Vol.\ 1
\bibitem[1]{1}
Wambsganss, J., Paczy\'nski, B. \& Schneider, P.\ 1990, ApJ 358, L33
\bibitem[1]{1}
White, M., van Waerbeke, L. \& Mackey, J.\ 2002, ApJ 575, 640
\bibitem[1]{1}
Williams, R.E., Blacker, B., Dickinson, M.\ et al.\ 1996, AJ 112, 1335
\bibitem[1]{1}
Wilson, G., Kaiser, N. \& Luppino, G.A.\ 2001, ApJ 556, 601
\bibitem[2000]{Wittm00}
  Wittman, D.M., Tyson, J.A., Kirkman, D., Dell'Antonio, I. \&
Bernstein, G. 2000, Nat, 405, 143
\bibitem[1]{1}
Wittman, D., Tyson, J.A., Margoniner, V.E., Cohen, J.G. \&
Dell'Antonio, I.P.\ 2001, ApJ 557, L89
\bibitem[1]{1}
Wittman, D., Margoniner, V.E., Tyson, J.A., Cohen, J.G. \&
Dell'Antonio, I.P.\ 2002, astro-ph/0210120
\bibitem[1]{1}
Wittman, D.\ 2002, in Lecture Notes in Physics 608, Gravitational
Lensing: An Astrophysical Tool, ed. F.\ Courbin \& D.\ Minniti
(Berlin: Springer), 55
\bibitem[1]{1}
Zaldarriaga, M. \& Scoccimarro, R.\ 2002, astro-ph/0208075

\end{thebibliography}
\end{document}